\documentclass[sigplan,10pt]{acmart}

\renewcommand\footnotetextcopyrightpermission[1]{}
\setcopyright{none}
\usepackage{hyperref}
\usepackage{url}

\usepackage{graphicx}
\usepackage{algorithm}
\usepackage{algpseudocode}
\usepackage{tabularx}
\usepackage[labelfont={small,it,bf},textfont={small,it}]{caption}
\usepackage{subcaption}
\usepackage{svg}
\usepackage{wrapfig}
\usepackage{multirow}
\usepackage[normalem]{ulem}
\usepackage[utf8]{inputenc} 
\usepackage[T1]{fontenc}    
\usepackage{hyperref}       
\usepackage{url}            
\usepackage{booktabs}       
\usepackage{amsfonts}       
\usepackage{nicefrac}       
\usepackage{microtype}      
\usepackage{xcolor}         
\usepackage{amsmath}
\usepackage{xspace}
\usepackage{paralist}
\usepackage{tikz}
\usepackage[compact,small]{titlesec}
\titlespacing{\section}{1pt}{2pt}{2pt}
\titlespacing{\subsection}{1pt}{\parskip}{\parskip}
\titlespacing{\subsubsection}{1pt}{\parskip}{-\parskip}

\usepackage[subtle]{savetrees}

\usepackage{tabularx}
\usepackage{dsfont}
\usepackage{enumitem}

\renewcommand{\paragraph}[1]{\vspace{0.03in}\noindent{\bf #1}}

\usepackage{listings}
\usepackage{xcolor}

\definecolor{codegreen}{rgb}{0,0.6,0}
\definecolor{codegray}{rgb}{0.5,0.5,0.5}
\definecolor{codepurple}{rgb}{0.58,0,0.82}
\definecolor{codered}{rgb}{0.6,0,0}

\lstdefinestyle{mystyle}{
    language=Python,                
    backgroundcolor=\color{white}, 
    commentstyle=\color{codered},  
    keywordstyle=\color{codegreen}\bfseries,
    numberstyle=\tiny\color{codegray},
    stringstyle=\color{codepurple},
    basicstyle=\ttfamily\scriptsize, 
    breakatwhitespace=false,        
    breaklines=true,    
    captionpos=t,     
    keepspaces=true,                 
    numbers=left,          
    numbersep=5pt,          
    showspaces=false,                
    showstringspaces=false,
    showtabs=false,                  
    tabsize=2,              
    frame=single,           
    emph={prism3d}, 
    emphstyle=\color{magenta}\bfseries
}
\definecolor{darkgreen}{rgb}{0.0,0.5,0.0}

\newcommand{\workload}[0]{Point-based Differentiable Rendering\xspace}
\newcommand{\name}[0]{Gaian\xspace} 
\newcommand{\workloadabbrev}[0]{PBDR\xspace}

\newcommand{\comments}[1]{}
\renewcommand{\comments}[1]{#1}

\newcommand{\xw}[1]{\comments{\textcolor{orange}{\textbf{XW:} #1}}}

\newcommand*\whitecircle[1]{\tikz[baseline=(char.base)]{\node[shape=circle, draw=black, fill=white, minimum size=10pt, inner sep=0pt] (char) {#1};}}

\begin{document}
\pagestyle{plain}
\title{Scaling Point-based Differentiable Rendering for Large-scale Reconstruction} 

\author{Hexu Zhao}
\affiliation{
  \institution{New York University}
  \city{New York}
  \state{NY}
  \country{USA}
}
\email{hz3496@nyu.edu}

\author{Xiaoteng Liu}
\affiliation{
  \institution{New York University}
  \city{New York}
  \state{NY}
  \country{USA}
}
\email{xiaoteng.liu@nyu.edu}

\author{Xiwen Min}
\affiliation{
  \institution{New York University}
  \city{New York}
  \state{NY}
  \country{USA}
}
\email{xm2336@nyu.edu}

\author{Jianhao Huang}
\affiliation{
  \institution{New York University}
  \city{New York}
  \state{NY}
  \country{USA}
}
\email{jh9885@nyu.edu}

\author{Youming Deng}
\affiliation{
  \institution{Cornell University}
  \city{Ithaca}
  \state{NY}
  \country{USA}
}
\email{ymdeng@cs.cornell.edu}

\author{Yanfei Li}
\affiliation{
  \institution{Pacific Northwest National Laboratory (PNNL)}
  \city{Richland}
  \state{WA}
  \country{USA}
}
\email{yanfei.li@pnnl.gov}

\author{Ang Li}
\affiliation{
  \institution{Pacific Northwest National Laboratory (PNNL)}
  \city{Richland}
  \state{WA}
  \country{USA}
}
\affiliation{
  \institution{University of Washington}
  \city{Seattle}
  \state{WA}
  \country{USA}
}
\email{ang.li@pnnl.gov}

\author{Jinyang Li}
\affiliation{
  \institution{New York University}
  \city{New York}
  \state{NY}
  \country{USA}
}
\email{jinyang@cs.nyu.edu}

\author{Aurojit Panda}
\affiliation{
  \institution{New York University}
  \city{New York}
  \state{NY}
  \country{USA}
}
\email{apanda@cs.nyu.edu}


\begin{abstract}

Point-based Differentiable Rendering (PBDR) enables high-fidelity 3D scene reconstruction, but scaling PBDR to high-resolution and large scenes requires efficient distributed training systems.
Existing systems are tightly coupled to a specific PBDR method. 
And they suffer from severe communication overhead due to poor data locality.
In this paper, we present \name, a general distributed training system for PBDR. 
\name provides a unified API expressive enough to support existing PBDR methods, while exposing rich data-access information, which \name leverages to optimize locality and reduce communication. 
We evaluated \name by implementing 4 PBDR
algorithms. Our implementations achieve high performance and resource
efficiency: across six datasets and up to 128 GPUs, it reduces communication by up to 91\% and 
improves training throughput by $1.50$---$3.71\times$. 


\end{abstract}


\settopmatter{printfolios=true}
\maketitle
















\section{Introduction}
\label{sec:intro}


\workload (\workloadabbrev) methods \cite{3dgs, 2dgs, 3dcs, survey, 3dgssurvey, metasapiens} have recently achieved remarkable breakthroughs in complex scene reconstruction and photorealistic novel view synthesis, which are long-standing challenging tasks in computer graphics.
Unlike traditional rendering methods, this new approach models a 
3D scene using a large collection of parameterized ``points'' (e.g. 3D Gaussian \cite{3dgs}) and a differentiable rendering pipeline.  The parameters of each point specifies its 3D position, color, opacity etc and are {\em learned} through mini-batch gradient descent to achieve accurate reconstruction.
\workloadabbrev methods have been shown to outperform previous neural rendering approach such as NeRF \cite{nerf, mipnerf} in terms of both rendering speed and rendered image fidelity. 
Beyond traditional graphics applications in gaming and AR/VR\cite{vr-gs,3dgs4vr}, \workloadabbrev has also been essential to build photo-realistic simulators for training autonomous driving and embodied AI \cite{nvidia_nurec,vid2sim,realtosim3dgs}, and serve as a foundational representation for World Models \cite{worldlabs_marble,GWM}.

Training a \workloadabbrev model requires a dataset of images with diverse camera views that capture the scene from different positions and orientations. 
Each training iteration renders a batch of images according to their camera views and the current point parameters, compares the rendered results with the actual images, and updates the parameters using the calculated gradients. In order to accurately reconstruct a large 3D scene at high-resolution~\cite{google_street, google_earth, hierarchicalgaussians, citygaussian, citygaussianv2}, one must correspondingly scale the size of the \workloadabbrev model, aka the number of points. Large models in practice can reach hundreds of millions or even billions of points. The resulting large model size and computational load make it impractical to train on a single GPU---or even a single node---within reasonable time and memory constraints. Therefore, training must be distributed across multiple GPUs.  

\begin{figure}[t] 
    \centering
    \includegraphics[width=1\linewidth]{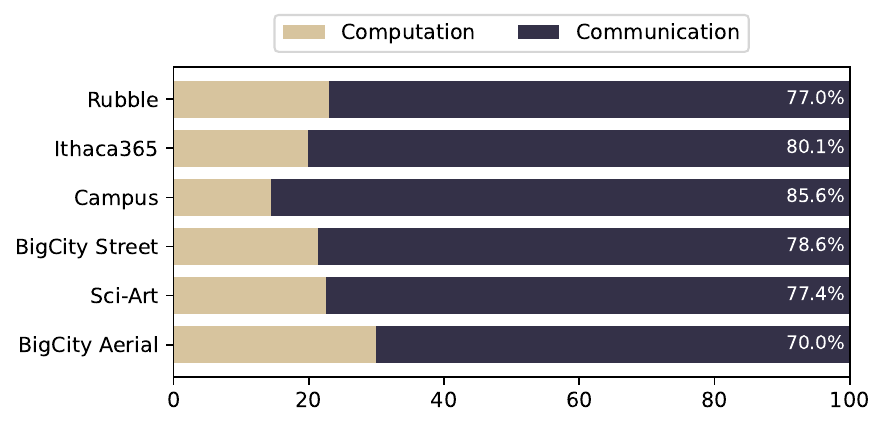} 
    \caption{ 
    Communication time as a fraction of each training step time in an existing system for distributed 3D Gaussian Splatting~\cite{gsplat}. 
    Y-axis lists the dataset with its experiment setting given by Table \ref{tab:scenes-main}. 
    }
    \label{fig:communication-challenge}
\end{figure}

Recently, there have been a few distributed \workloadabbrev training systems~\cite{grendel, gsplat}.  However, they have two major limitations.  First, existing systems are  specialized to a single \workloadabbrev algorithm, namely, the very popular 3D Gaussian Splatting~\cite{3dgs}.  However, as developing better \workloadabbrev algorithms is a very active research area, a distributed training system must be general to support different algorithms. Second, existing systems randomly partition the model state (aka points) across GPUs (thus addressing the memory capacity limit), and randomly assign different GPUs to render each image in a training batch.  Unfortunately, this strategy requires significant data movement, leading to high
communication costs: during training, each GPU tends to access points stored across all GPUs
in order to render its images.  As shown in Figure~\ref{fig:communication-challenge}, communication time accounts for 70–85\% of each training step across different datasets in an existing system~\cite{grendel}.

In this paper, we propose a distributed training system, \name 
.
It provides a common interface that hides the internal differences between \workloadabbrev algorithms while exposing those aspects necessary for distribution. Thus, once developers implement a \workloadabbrev algorithm using \name's interface, the distribution of training can be done transparently.   

\name partitions points and the rendering of images across GPUs.  However, rather than partitioning them randomly like existing systems~\cite{grendel,gsplat}, \name optimizes the placement of points and image rendering to exploit locality of access.  Our setting for locality optimization differs from other types of distributed computation (e.g. big data analytics, distributed GNN and recommendation) in important ways (\S\ref{sec:related}): 1) \name is explicitly aware of \workloadabbrev's access pattern via its programming API, and 2) there exist two forms of spatial locality: locality among points and locality between cameras and their viewable points. The former remains stable over a relatively long time period.  To leverage these insights,  
we develop a two-stage optimization strategy that combines offline and online optimization. The offline component optimizes the placement of points according to the access pattern of images to points over the entire dataset. Points likely to be access together are assigned to the same GPU to minimize communication. The online component that determines how to assign images in a training batch to GPUs based the access pattern of images within a given batch.  As off-the-shelf optimizers are too slow for online optimization, we propose an optimizer based on Linear Sum Assignment~\cite{lsa} and iterative local search~\cite{stochasticlocalsearch}.



We build \name and evaluate its performance using six different datasets for up to $128$ GPUs.
Our experiments show that \name outperforms the existing system (gsplat~\cite{gsplat}) by 1.50–3.71$\times$.
\name's locality-aware distribution reduces communication volume by up to 91\%. When scaling from 8 to 128 GPUs, \name attains $12.7\times$ speedup.  To the best of our knowledge, we manage to train the largest \workloadabbrev point cloud with 500 million points (29.5 billion parameters), achieving state-of-the-art reconstruction quality on the MatrixCity dataset~\cite{matrixcity}.  Finally, we also implement and evaluate a much more complex \workloadabbrev algorithm for 3D video reconstruction. 

In summary, this paper makes the following contributions:
\begin{compactitem}
\item We propose a set of general and easy-to-use programming APIs to support distributed training of a variety of \workloadabbrev algorithms.  Our APIs hide algorithmic details from the underlying distribution system to achieve generality while also expose data access pattern to enable locality optimization. 
\item We build \name to support distributed \workloadabbrev training.  We develop a two-stage procedure 
to optimize locality of access. It consists of offline optimization for the partitioning of model state (aka points) and online optimization for assigning compute (aka image rendering) to GPUs at each training iteration.
\item We built and evaluated four \workloadabbrev algorithms including one that performs complex 3D video reconstruction.  The variety of algorithms demonstrate the generality of \name's APIs. Our large-scale evaluations up to 128 GPUs show that \name achieves good training throughput and scalability compared to existing systems.
\item Using \name, we train the largest 3D Gaussian splatting model to date with up to 500 million points and achieve state-of-the-art reconstruction quality.
\end{compactitem}

\section{Background and Motivation}
\label{sec:background}

\subsection{\workload (\workloadabbrev)} 
\label{sec:background:workloads-description}

\begin{figure}
    \centering
    \includegraphics[width=\linewidth]{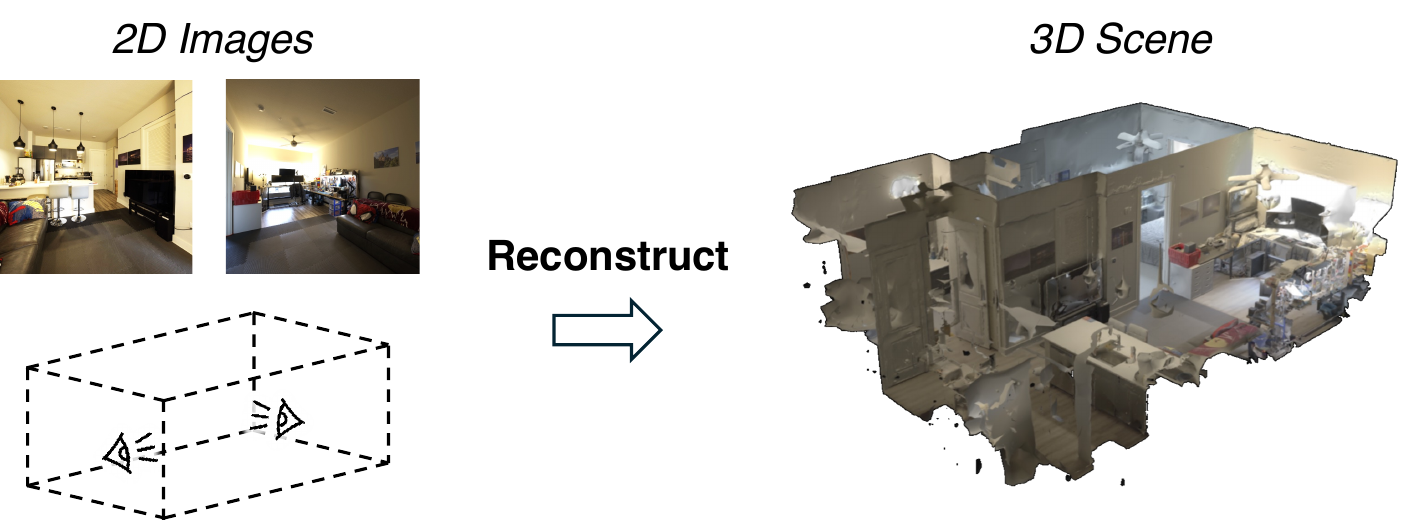}
    \caption{\workloadabbrev algorithms are used to reconstruct complex 3D scenes using 2D images with different camera views. This example is based a dataset in~\cite{eyeful}.} 
    \label{fig:room-reconstruction}
\end{figure}

\workload methods are used to reconstruct complex scenes and synthesize novel views using a collection of images captured from diverse camera views of the scene. It has wide applications ranging from AR/VR~\cite{eyeful, 3dgs4vr}, simulator for robotics and autonomous systems~\cite{nvidia_nurec,vid2sim,realtosim3dgs} and World Models~\cite{worldlabs_marble,GWM}.
Figure~\ref{fig:room-reconstruction} shows an example of reconstructing a 3D model of the living room using multiple 2D images from different camera views.  Given the room's 3D model, users can rotate, zoom into, and generate a photorealistic image from a new viewpoint never previously captured.
Beyond static scenes, some \workloadabbrev algorithms \cite{yang4dgs, dynamic1, freetimegs} can also reconstruct dynamic scene that evolves over time (aka, 3D Video). 


\paragraph{\workloadabbrev models and their rendering. } 
In \workloadabbrev methods, the scene is modeled as a learnable ``point cloud''
which is a (potentially very large) collection of points whose parameters can be optimized through training. Each ``point'', which is actually a 3D ellipsoid, has several (parameterized) attributes that attempt to capture geometry and appearance smoothly.  The first \workloadabbrev algorithm, 3D Gaussian Splatting (3DGS)~\cite{3dgs} uses 3D Gaussian ellipsoid with five attributes: 3D position, scales, rotation, opacity, and color-related spherical harmonics.  Since then, a large number \workloadabbrev algorithms have been proposed with different point types, including 2D Gaussian~\cite{2dgs}, 3D Convex~\cite{3dcs}, Triangle Mesh~\cite{trianglesplatting}, 
Generalized Exponential Kernel~\cite{gexpsplat} and many others~\cite{3dgssurvey}. 


To render an image given a model and a camera view, a \workloadabbrev algorithm proceeds through several steps.  First, as a performance optimization, it discards points that lie outside of the camera's field of view (frustum). Then, it projects each of the remaining points into the camera's 2D image plane as a ``splat''. Lastly, it sorts these view-dependent splats by their distance to the camera plane, and the renderer computes the color of each pixel front to back, blending the overlapping splats until full opacity is reached.



\paragraph{Training a \workloadabbrev model.} 
The rendering of a \workloadabbrev algorithm is differentiable. This allows point attributes (e.g., position) to be optimized via gradient descent by calculating the loss between the rendered image and the ground-truth image for the image's camera view. 
The training process goes as follows. First, we initialize each point using an initial point cloud based on COLMAP~\cite{colmap} or LiDAR. Then, training proceeds iteratively: at each step, we randomly select a batch of images from the training dataset and render according to their camera views using the current point cloud. The rendered images are then compared to the corresponding training images to compute the loss. Finally, backpropagation is performed to calculate the gradients that are used to update the point cloud parameters for the next iteration. This standard mini-batch SGD training process is augmented with periodic densification, which increases the density of the point cloud by adding more points to the model. The training is repeated until convergence. 



\subsection{Limitations of existing work and our approach} 
\label{sec:background:characteristics}

Existing works~\cite{grendel,gsplat} support distributed training for 3D Gaussian Splatting~\cite{3dgs}, the first and most popular \workloadabbrev algorithm.  
These systems partition the point cloud across GPUs.  Given a batch of training images, each GPU performs point culling and splatting for all camera views in the batch using its local partition of points. Each GPU is in charge of rendering a subset of images in the batch. As some in-frustum points needed for rendering an image on a GPU may reside on other GPUs, view-dependent splats must be shuffled across GPUs in an all-to-all pattern before rendering. Existing systems have two major limitations.

\paragraph{Limitation \#1: Lack of programmable support for different \workloadabbrev algorithms.} The first major limitation of existing systems is that they are custom built for one particular \workloadabbrev algorithm, 3DGS.  However, the rapid growth of research on \workloadabbrev has produced many algorithms~\cite{2dgs,3dcs,trianglesplatting,gexpsplat, betasplat,3dhalfsplat,linearsplat} and more will undoubtedly show up in the future. Thus, it is imperative for a distributed training system to provide general support of the whole class of \workloadabbrev algorithms instead of only 3DGS.  

We address this challenge by building \name as a programmable training system that can be customized by users to support different \workloadabbrev algorithms.  As all \workloadabbrev proceed through 
the same three key steps to cull, splat and render points, \name provides three corresponding user-defined functions to allow full customization of these three steps.  This allows the separation of the underlying distributed training system from the details of specific PBDR algorithm it supports.
Furthermore, \name's interface is designed to make explicit the algorithm’s data access pattern, aka which parts of the point cloud are accessed by each camera view, which we can exploit to improve locality and reduce communication during training.

\paragraph{Limitation \#2: Lack of locality optimization for data access.} Existing systems randomly distribute points and image rendering to GPUs, resulting in heavy communication as in-frustum points reside on remote GPUs and their splats must be fetched for rendering. As seen in Figure~\ref{fig:communication-challenge}, the time spent doing communication can exceed the compute time.

\begin{figure}[t]
    \centering
    \begin{subfigure}{0.48\linewidth}
        \includegraphics[width=\linewidth]{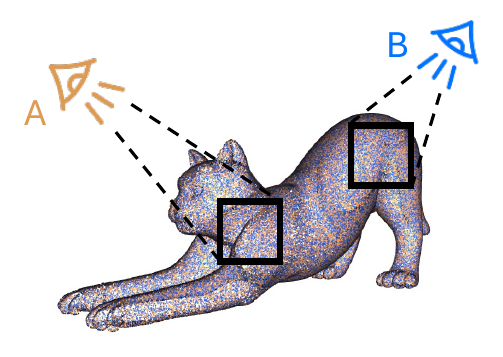}
        \caption{Random Distribution}
        \label{fig:random-distribution}
    \end{subfigure}
    \hfill
    \begin{subfigure}{0.48\linewidth}
        \includegraphics[width=\linewidth]{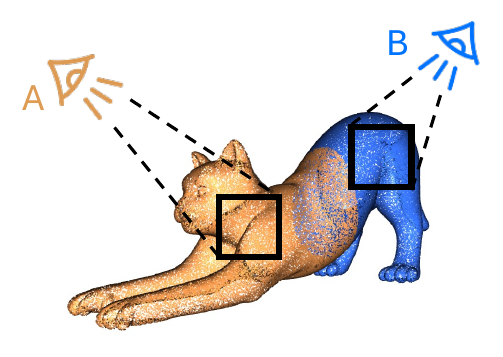}
        \caption{Locality-aware Distribution}
        \label{fig:locality-aware-distribution}
    \end{subfigure}
    \caption{
    Random distribution vs. locality-aware distribution of model state and compute. 
    }
    \label{fig:compare-random-localityaware}
\end{figure}

Although optimizing for locality is a common theme among all distributed computation, \workloadabbrev presents a unique workload that warrants special care.  In particular, our design exploits spatial locality using a two-stage procedure that combines online and offline optimization.  (1) Instead of randomly assign points to GPUs, we assign points that are likely to be accessed together to the same GPU. This step is done offline since the optimality of the assignment does not change much during training. 
(2) Instead of randomly assigning camera views to GPUs, we prefer to co-locate the rendering of an image to the GPU storing the most points needed by the image's view.  However, since the batch of images for a training iteration is randomly sampled, greedily assigning each view to its most preferred GPU can cause severe load imbalance. In order to balance load and reduce communication, we propose a speedy online optimization procedure to assign images in a batch to GPUs. Figure \ref{fig:compare-random-localityaware} gives an example to illustrate the benefit of \name's locality-aware assignment.  The points are colored according to which GPU it has been assigned to. As shown in Figure~\ref{fig:compare-random-localityaware}(b), \name's point assignment takes advantage of spatial locality, and the rendering of view $A$ (or $B$) is assigned to the GPU holding most of the points needed by $A$ (or $B$), resulting in drastic communication reduction compared to random distribution (Figure~\ref{fig:compare-random-localityaware}(a)).

\section{Programming \workloadabbrev algorithms on \name} 
\label{sec:overview}

This section discusses the programming abstractions (\S\ref{subsec:api}) and provide example \workloadabbrev implementations (\S\ref{subsec:examples}). 



\begin{figure}[h] 
    \centering
    \includegraphics[width=\linewidth]{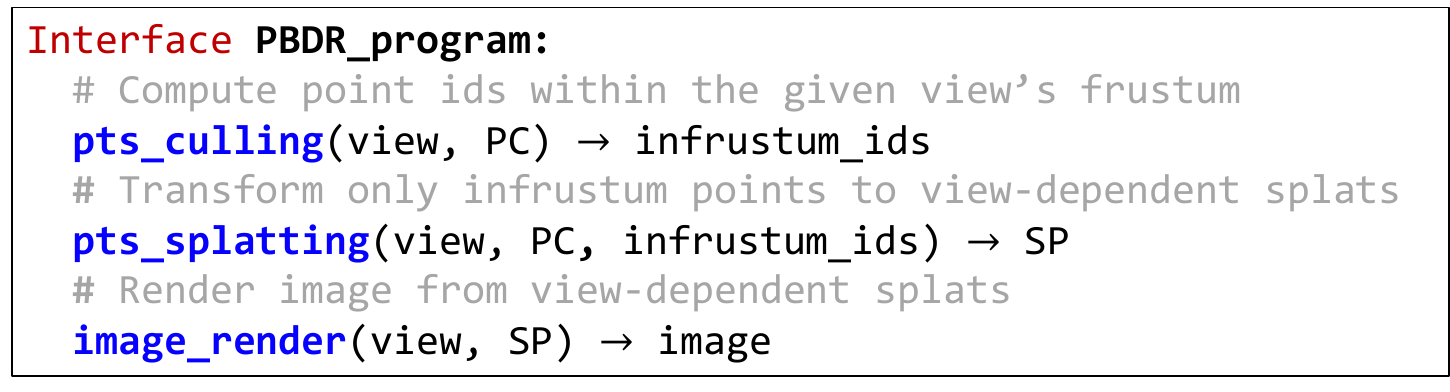} 
    \caption{\name abstracts a PBDR algorithm using three user-defined functions. } 
    \label{fig:sys-api}
\end{figure}

\subsection{Abstracting a \workloadabbrev algorithm} 
\label{subsec:api} 

\name provides a common interface which an algorithm developer must use to structure their code to enable distribution. 
The interface design must strike a fine balance between two considerations: (1) it must be general enough to support a wide range of PBDR algorithms without revealing much of their algorithmic internals to the underlying training system. (2) it should also expose sufficient information for the underlying training system to perform distribution and to optimize for the locality of access.

Our interface exposes every \name algorithm's model state, aka the point cloud, which is represented as a dictionary of tensors that store point attributes.  Each tensor has the shape $S\times l$ where $S$ is the number of total points and $l$ is number of floating point needed to store this attribute. For example, given point cloud state $Points$, the 3D Gaussian Splatting~\cite{3dgs} algorithm can use tensor $Points.xyz$ (of shape $S\times 3$) to store each point's positions in 3D space and another tensor $Points.opacity$ (of shape $S\times 1$) to store each point's opacity.  After in-frustum points are projected to 2D camera plane, the resulting view-dependent splats are also represented as a dictionary of tensors, where each tensor's first dimension equals to the number of in-frustum points.

\begin{figure}
    \centering
    \includegraphics[width=\linewidth]{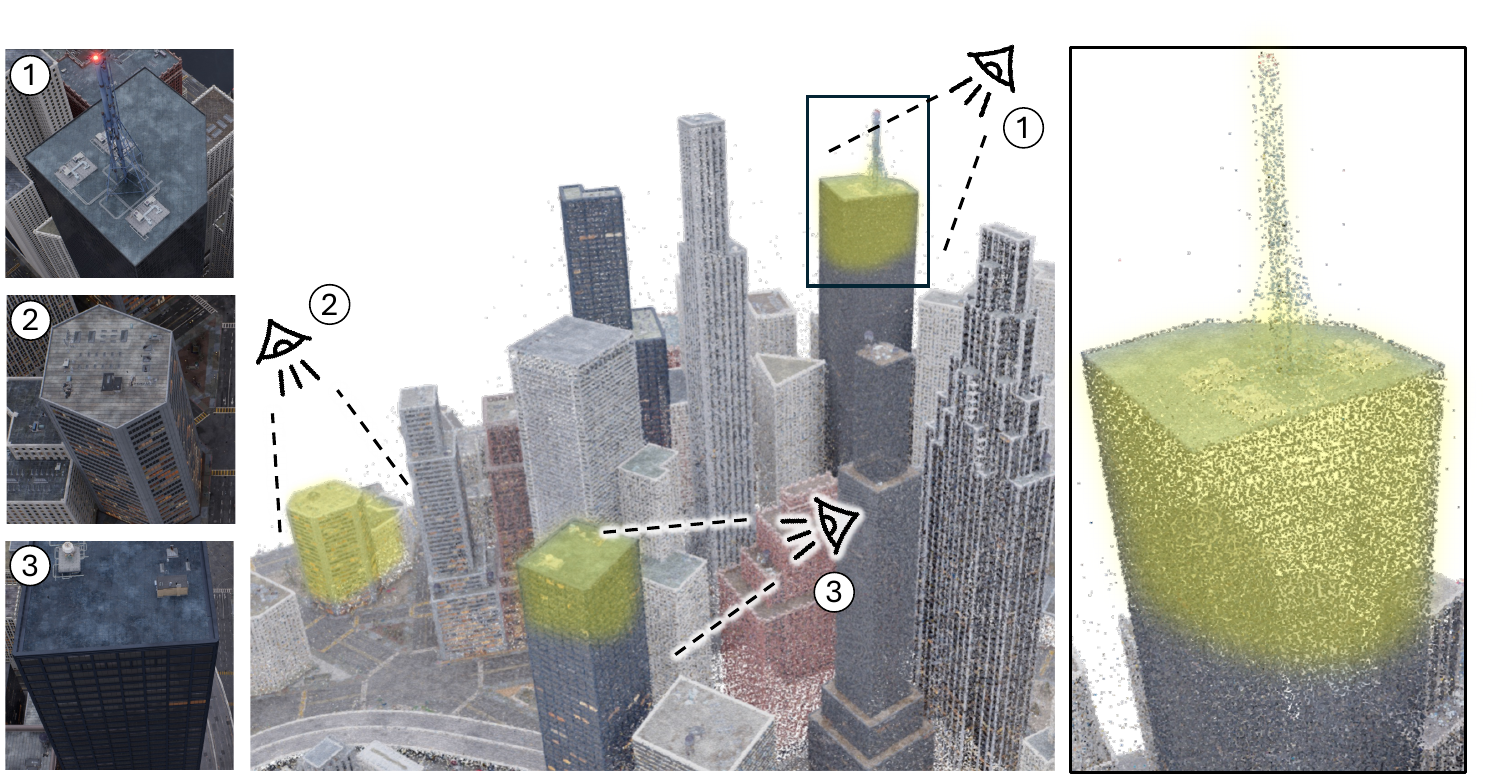}
    \caption{Point culling selects the subset of points that fall within the camera's view frustum. The middle shows the point cloud of MatrixCity scene~\cite{matrixcity}.  \protect\whitecircle{1}, \protect\whitecircle{2}, and \protect\whitecircle{3} denote different camera views and their in-frustum points, shown in yellow, respectively. The right panel zooms in to visualize the in-frustum points for view \protect\whitecircle{1}.} 
    \label{fig:infrustum}
\end{figure}

\name's interface abstracts a \workloadabbrev algorithm into three user-provided functions: 
\textcolor{blue}{{\sf pts\_culling}}, \textcolor{blue}{{\sf pts\_splatting}} and \textcolor{blue}{{\sf image\_render}}, 
as shown in Figure~\ref{fig:sys-api}.  The \textcolor{blue}{{\sf pts\_culling}} function takes as arguments a specific camera view ({\sf v}) as well as (a shard of) the point cloud state ({\sf PC}), and returns a 1D tensor containing the indices of the points that remain after culling. 
The standard implementation of \textcolor{blue}{{\sf pts\_culling}},  calculates the camera view's frustum and culls those points outside the frustum, as shown by the example in Figure~\ref{fig:infrustum}.  However, some algorithms require non-standard \textcolor{blue}{{\sf pts\_culling}}. For example, 3D video reconstruction algorithms cull both spatial out-of-frustum points and temporal non-present points at the camera’s capture time. The \textcolor{blue}{{\sf pts\_splatting}} function projects each 3D point to the 2D camera plane, creating view-dependent splats, also represented as a dictionary of tensors ({\sf SP}).  Since such processing only needs to be done on in-frustum points, it takes as the argument {\sf infrustum\_ids}, which has previously been returned by \textcolor{blue}{{\sf pts\_culling}}.  The \textcolor{blue}{{\sf image\_render}} function takes as input the camera view and {\em all} the view-dependent splats ({\sf SP}), and renders the resulting image.  
The functions \textcolor{blue}{{\sf pts\_splatting}} and \textcolor{blue}{{\sf image\_render}} should be differentiable. The differentiation is usually done automatically by the underlying ML framework. 

We note that \name's interface exposes a \workloadabbrev algorithm's data access pattern from images to points via \textcolor{blue}{{\sf pts\_culling}}.  This gives the underlying distribution system valuable information to fully optimize for access locality.

\subsection{Example Algorithm Implementations}
\label{subsec:examples}

\begin{figure}[h] 
    \centering
    \includegraphics[width=\linewidth]{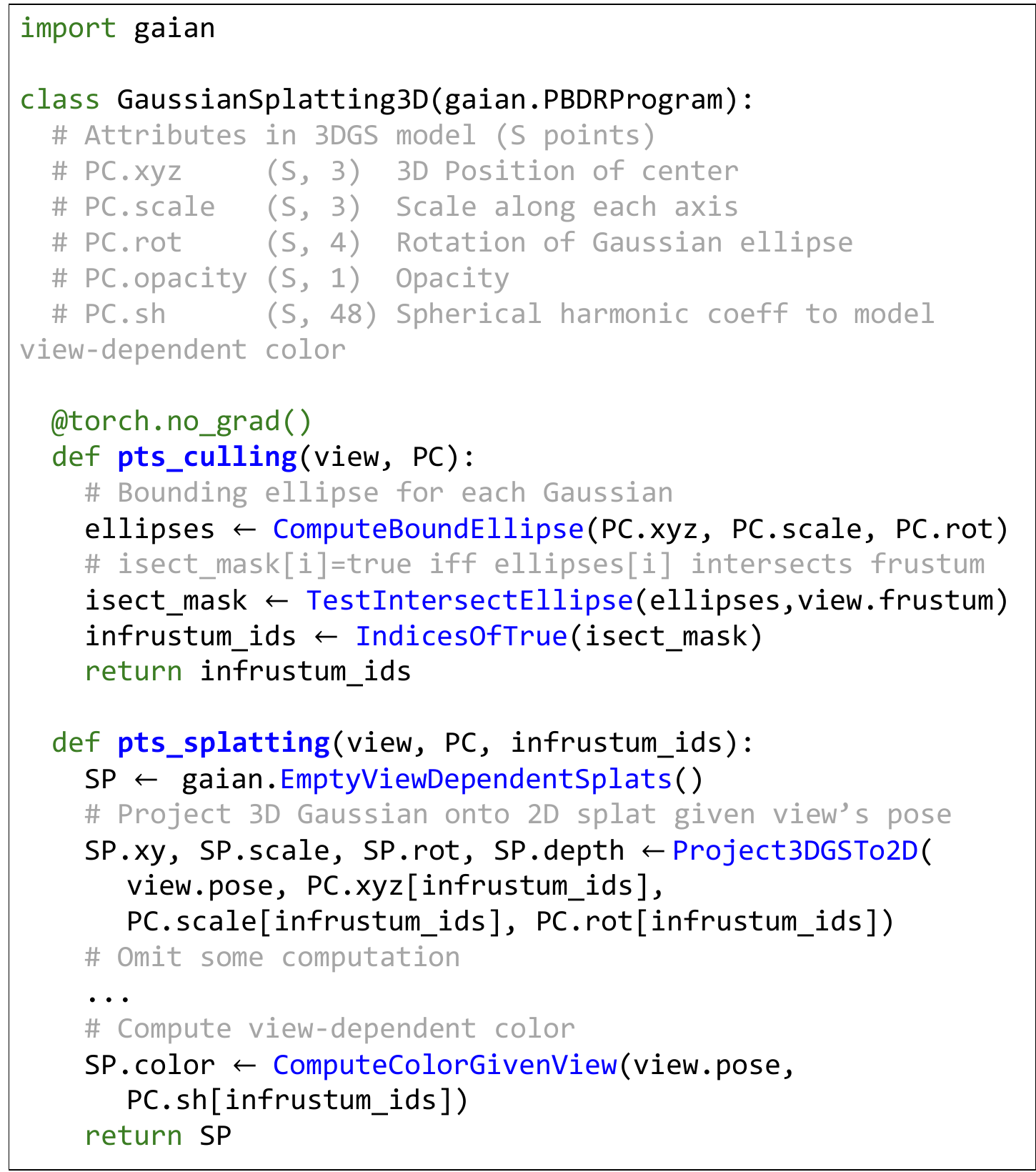} 
    \caption{Implement 3DGS with \name API.  }
    \label{fig:3dgs-implement-api}
\end{figure}

We provide an example implementation of the 3DGS algorithm~\cite{3dgs} built using \name's abstractions.  Additional more complex examples including 3D Convex Splatting (3DCX) and 4D Gaussian Splatting for 3D video, can be found in Appendix~\ref{sec:appendix:additional-examples}. 


\paragraph{3DGS.} 
As shown in Figure~\ref{fig:3dgs-implement-api}, 3DGS represents each point as an anisotropic 3D Gaussian whose point attributes include 3D position, per-axis scale, and rotation, opacity and spherical-harmonics for color. 
For \textcolor{blue}{{\sf pts\_culling}}, each Gaussian is approximated by a 3D bounding ellipse; Gaussians whose ellipses fall completely outside the camera frustum ({\sf view.frustum}) are not returned. 
For \textcolor{blue}{{\sf pts\_splatting}}, each in-frustum Gaussian is projected onto the image plane given the camera pose ({\sf view.pose}). The resulting view-dependent attributes including color, scale, and depth, collectively form the intermediate per-view splat state {\sf SP} and returned by the function. We omitted the implementation of {\sf image\_render} for brevity. 

Different \workloadabbrev algorithms have varying point attributes as well as varying implementations of \textcolor{blue}{\sf pts\_cull} and \textcolor{blue}{\sf pts\_splatting} (see additional examples in Appendix~\ref{sec:appendix:additional-examples}).  Even for the 3DGS, there are several interesting variants.  For example, one can replace the bounding ellipse with a bounding sphere for a simpler implementation at the cost of more false positives.  Conversely, users can calculate a tighter bounding ellipse according to~\cite{speedysplat}, which restricts the bound to the higher-density region of each Gaussian to reduce the number of in-frustum points.  These example variants can be supported easily by modifying \textcolor{blue}{\sf pts\_culling}.

{\small
\begin{algorithm}
\caption{\name's workflow for a training iteration,  from GPU $k$'s perspective} 
\label{alg:pbdr-algorithm}

\begin{algorithmic}[1]
    
\Statex \textbf{Input:} 
$\textit{PC}_k$ : point cloud shard stored on GPU $k$.
\Statex $batch$: batch of images for this training iteration

\Statex \rule{\linewidth}{0.4pt}
\Statex \textit{\textbf{Forward Pass:} }
\ForAll{$v$ in $batch$} 
    \State $infrustum\_ids[v]_k \gets \textcolor{blue}{\texttt{pts\_culling}}(v.view, \textit{PC}_k)$
    \State $C[v]_k\gets \texttt{len}(infrustum\_ids[v]_k)$
    
    \State $SP[v]_k \gets \textcolor{blue}{\texttt{pts\_splatting}}(v.view,PC_k,infrustum\_ids[v]_k)$
\EndFor


\Statex //$\mathcal{A}$ is a $|batch|\times N$ tensor capturing access pattern. 
\State $\mathcal{A} \gets$ all-gather $C[\cdot]_{k}$ from all GPUs $k=1\ldots N$
\State //Image assignment: GPU W[v] is to render v.
\State $W \gets \texttt{AssignImages}(\mathcal{A})$

\State $SP_k \gets $all-to-all transfer of $SP[v]_k$ at GPU $k$ to GPU $W[v]$
\Statex \quad \quad \quad for all $v$ in $batch$


\State $loss\_total \gets 0$
\ForAll{$v$ such that $W[v] = k$}
    \State $image \gets \textcolor{blue}{\texttt{image\_render}}(v.view, \textit{SP}[v]_k)$
    \State $loss_v \gets \texttt{loss\_fn}(image, v.image)$
    \State $loss\_total \gets loss\_total + loss_v$
\EndFor

\Statex \rule{\linewidth}{0.4pt}
\Statex \textit{\textbf{Backward Pass:} } 
\ForAll{$v$ such that $W[v] = k$}
    \State $G\_image \gets \texttt{loss\_fn.backward}(\texttt{loss}_{v})$
    \State $G\_SP[v]_k \gets \textcolor{blue}{\texttt{image\_render.backward}}(v, G\_image)$
\EndFor 


\State //reverse all-to-all transfer of $SP_k$'s gradients
\State $G\_SP_k$ $\gets$ all-to-all transfer of $G\_SP[v]_i$ at GPU $i$=$W[v]$ to GPU $k$ for all $v$ in batch


\State Initialize $\textit{G\_PC}_k \gets$ zero tensors
\ForAll{$v$ in batch}
    \State $\textit{G\_PC}_k \mathrel{+}$= \textcolor{blue}{\texttt{pts\_splatting.backward}}(
    \Statex \quad\quad\quad\quad\quad\quad\quad $v$, $G\_SP_k[v]$,\ $infrustum\_ids[v]_k$) 
\EndFor
\State //update point cloud parameters
\State $PC_k \gets \texttt{AdamUpdate}(PC_k, G\_PC_k)$ 
\end{algorithmic}
\end{algorithm}
}

\section{System Design} 
\label{sec:design} 

\begin{figure*}[t]
    \centering
    \includegraphics[width=0.85\linewidth]{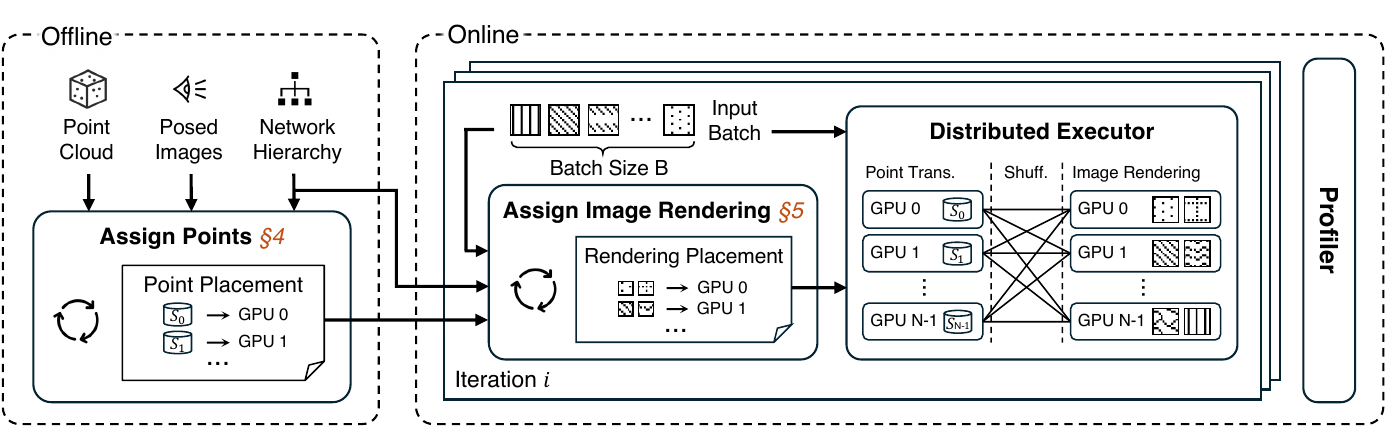} 
    \caption{System overview.
    The offline component assigns points to GPUs (\S\ref{subsubsec:pointplacement}). During training, the online optimizer assigns image rendering to GPUs (\S\ref{subsubsec:imageplacement}) based on data-compute locality. Distributed Executor handles the forward and backward pass compute and communication.
    }
    \label{fig:sys_architecture} 
\end{figure*}

\subsection{Architecture overview}

\name employs a hybrid parallelism approach that partitions both the model state (data parallelism) and the rendering computation (image/view parallelism) across GPUs. While existing systems~\cite{grendel} also adopt a similar distribution strategy, their implementation is tied to a single algorithm.  By contrast, as \name's programming abstractions expose \workloadabbrev algorithm's point cloud as well as the data access pattern from images to points, \name supports the distribution of all algorithms. Equally importantly, as the data access is made explicit, \name optimizes locality of access through a hybrid offline/online procedure. Figure~\ref{fig:sys_architecture} illustrates the overall system architecture. \name uses an offline component to assign points to GPUs and an online component that assigns each image in the current training iteration to be rendered by a GPU. 

Algorithm~\ref{alg:pbdr-algorithm} provides the pseudocode for how \name performs a single training iteration from the perspective of GPU $k$. Each GPU $k$ stores a shard of point clouds, as determined by the \name's offline assignment procedure, which remains stable across training iterations.  At the start of each iteration, all camera views of the current batch is sent to all GPUs.  In the forward pass (lines 1-15), GPU $k$ calculates the subset of in-frustum points among its local points ($PC_k$) and performs view-dependent point splatting for each view in the batch. The online rendering placement algorithm then assigns images to GPUs according to the calculated access pattern. For each assigned image $v$, GPU $k$ retrieves the in-frustum view-dependent splats $SP[v]_i$ from every other GPUs $i$, performs image rendering and calculates loss w.r.t. the corresponding ground truth image. The backward pass 
(lines 16-25) proceeds by reversing the computation and communication pattern.  Lastly, the point cloud parameters are updated using the gradients computed by the backward pass (lines 26-27).




\subsection{Locality Optimization}

\name's two-stage optimization consists of an offline component to optimize the partitioning of model state (\S\ref{subsubsec:pointplacement}) and an online component to optimize the placement of compute, aka image rendering (\S\ref{subsubsec:imageplacement}).

\subsubsection{Offline point placement.}   
\label{subsubsec:pointplacement} 
A straightforward strategy of assigning points to GPUs is to partition according to geometric proximity ~\cite{citygaussian,vastgaussian,retinags}, e.g. using space-partitioning data structures (e.g., KD-trees) or spatial clustering methods (e.g., k-means~\cite{kmean4pointcloud}).
However, doing so is suboptimal as it overlooks the fact that data access locality can be heavily affected by the camera view: points that are not close to each other might nevertheless be seen by the same camera view. 

\begin{figure}[h]
    \centering
    \begin{subfigure}{0.45\linewidth}
        \includegraphics[width=\linewidth]{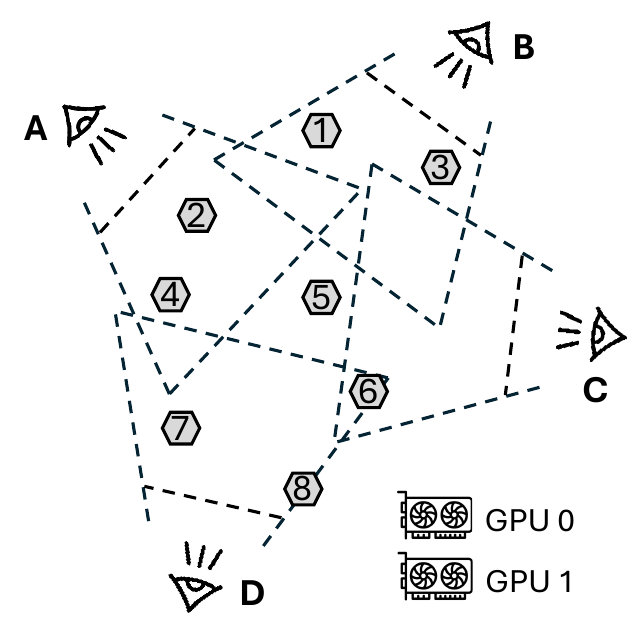}
        \caption{Camera Views}
        \label{fig:the-partition-scene}
    \end{subfigure}
    \hfill
    \begin{subfigure}{0.45\linewidth}
        \includegraphics[width=\linewidth]{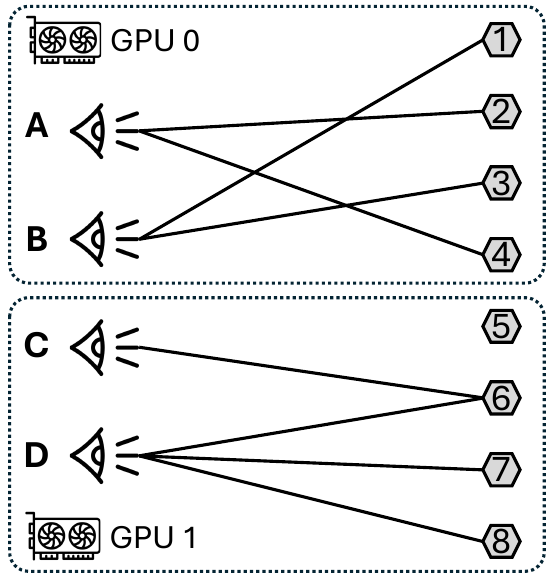}
        \caption{Partition a Bipartite Graph}
        \label{fig:bipartite-graph-partition}
    \end{subfigure}
    \caption{\name represents the access pattern of different camera views and points as a bipartite graph (a) and partitions the graph to assign points to GPUs (b).
}
    \label{fig:bipartite-partition-example} 
\end{figure} 
 

We represent the data access pattern explicitly as a bipartite graph and formulate the problem of point placement as a graph partitioning problem.  In the bipartite graph, an edge connects a point-$i$ to image-$j$ if point-$i$ is in image-$j$'s frustum, as shown in Figure~\ref{fig:the-partition-scene}. Graph partitioning assigns vertices into different partitions (aka GPUs), an example of which is shown in Figure~\ref{fig:bipartite-graph-partition}.  If an edge spans across different partitions, the point's splat must be communicated between GPUs in order to render the corresponding image.  Therefore, our objective is to minimize the number of cross-partition edges while keeping each part roughly equal in size (i.e., balanced memory usage across GPUs). To incorporate the latter load balancing constraint, we assign weights to vertices in the bipartite graph. An image vertex's weight is based on a rendering complexity heuristic---measured by the number of points accessed---to approximate the computation, memory, and communication cost needed to render an image. Graph partitioning is a well-studied problem in the literature \cite{graphpartition, graphpartitionsurvey, KLalgo, metis, spectralpartition,parmetis, PaToH, Zoltan}, allowing us to use the popular off-the-shelf METIS~\cite{metis} to generate partitions where each partition has relatively balanced node weights.  

We augment basic weighted graph partitioning in several ways to further improve performance.  First, assigning each point individually is very expensive when the model state is large (with hundreds of millions or even more points). We coarsen the assignment by grouping nearby points into a single placement unit. To do so, we sort all points in the point cloud using Z-order curves~\cite{zordercurve} that preserve spatial locality by ensuring that points close in 3D space remain close in the linear ordering. We then group each contiguous block of $G$ points along the ordering into a single point group, which is represented by a single point vertex in the bipartite graph.
We can adjust the placement granularity by adjusting $G$. More coarse-grained grouping makes it faster to compute point placement but compromises the placement quality.
In practice, we choose $G$ to be 1024---4096 to trade off speed with quality.  Second, cross-GPU communication is non-uniform with inter-node communication significantly slower than intra-node communication. To account for this, we perform partitioning hierarchically: first, we partition the point cloud across machines, assigning one partition to each machine. Then, we partition each machine's points among its local GPUs. 




\subsubsection{Online placement of image rendering.}
\label{subsubsec:imageplacement}

\begin{figure}[t]
    \centering
    \includegraphics[width=\linewidth]{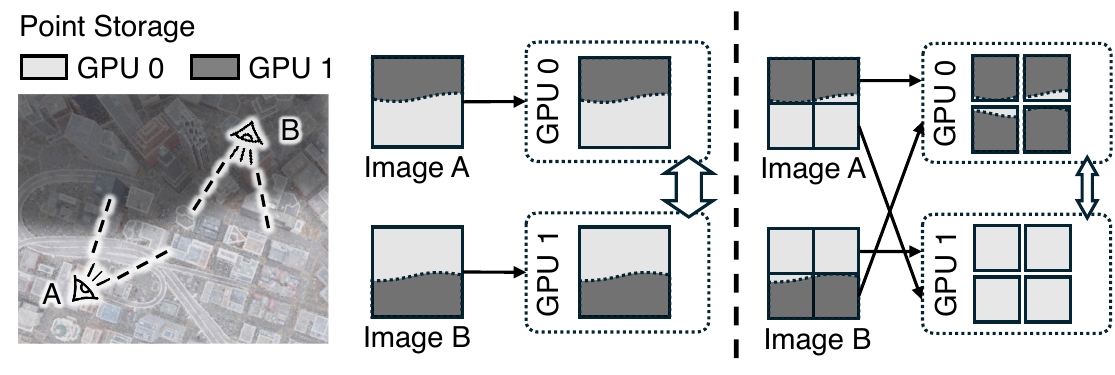}
    \caption{
    Whole image (the middle figure) vs. image patch placement (the right figure).
    } 
    \label{fig:design_pixelPatch}
\end{figure}

Although offline graph partitioning clusters both points and images, we only use the partitioning results to assign points and not compute.  This is because the bipartite graph is constructed for all images in the dataset while each training iteration must assign the images to render at different GPUs for a randomly chosen small batch of images.  Therefore, \name determines how to assign image rendering online during each iteration.
At each training iteration, every GPU is given the entire batch of $B$ image patches as input. Each GPU first performs frustum culling to determine the subset of points among those stored by its local memory for rendering each image patch (Algorithm~\ref{alg:pbdr-algorithm}, line 2). All GPUs exchange this information to obtain the patch-GPU access pattern matrix, $\mathcal{A}$ (Algorithm~\ref{alg:pbdr-algorithm}, line 6), where $\mathcal{A}_{j,k}$ captures the number of points needed from GPU $k$ in order to render image patch $j$.

We formulate placement optimization as Integer Linear Programming (ILP). Let $W$ be the placement solution where $W_j$ denotes the GPU assigned to render image patch $j$.  Our goal is to find the optimal $W$ with low communication and balanced rendering loads, with the following ILP formulation:

\begin{subequations}
\label{eq:optimization-formula}
\begin{align}
\text{minimize} \quad 
& \alpha \cdot \left(- \sum_{j=1}^B \mathcal{A}_{j, W_{j}} \right) \label{eq:total_comm} \\
& + \beta \cdot \max_k (\texttt{send}_k) + \gamma \cdot \max_k (\texttt{recv}_k) \label{eq:max_send_recv} \\
& + \delta \cdot \max_k (\texttt{comp}_k) \label{eq:render_max} \\
\text{subject to} \quad 
& \sum_{j=1}^{B}{\left[W_j=k\right]}= \frac{B}{N}, \quad \forall k \in \{1, \dots, N\}. \label{eq:balance_constraint}
\end{align}
\end{subequations}

  
    


In the above formulation, we combine the different optimization objectives linearly using configurable coefficients $\alpha, \beta, \gamma, \delta$ to trade off competing objectives. Objective (\ref{eq:total_comm}) aims to minimize total communication to render the batch of $B$ images.  Objective(\ref{eq:max_send_recv}) aims to minimize the imbalance of send and receive load, where $\texttt{send}_k= \sum_{j=1}^B({\left[W_j\neq k\right]\cdot \mathcal{A}_{j,k}})$\footnote{$\left[..\right]$ is an indicator function returning 1 if the predicate inside is true and zero otherwise} and $\texttt{recv}_k= \sum_{j=1}^B({\left[W_j=k\right] \cdot (\sum_{i=1}^{N}{\mathcal{A}_{j,i}}}-\mathcal{A}_{j,k}$)) capture GPU-$k$'s send and receive load, respectively. Lastly, objective(\ref{eq:render_max}) aims to minimize the imbalance of the rendering compute load, where $\texttt{comp}_k=\sum_{j=1}^B{(\left[W_j = k\right] \cdot \sum_{i=1}^{N}{\mathcal{A}_{j,i}})}$.  We also add a set of constraints(\ref{eq:balance_constraint}) to ensure that each GPU is assigned exactly the same number ($\frac{B}{N}$) of images. This is necessary to simplify system development using MPI.

\paragraph{A fast optimizer based on Linear Sum Assignment.}
ILP can be solved using standard solvers such as Gurobi~\cite{gurobi}. However, these general solvers are slow and have unpredictable running time, making them unsuitable for online optimization.
We develop a fast heuristic online solver using Linear Sum Assignment (LSA) with local search~\cite{stochasticlocalsearch}. Our solver works in two stages.
First, we apply LSA to obtain an initial placement $\mathcal{W}^{init}$ that satisfies the ILP constraints~\ref{eq:balance_constraint} and has low total communication volume. Then, we apply local search to incrementally adjust $\mathcal{W}^{init}$ to account for the load-balancing objectives.
\begin{itemize}[topsep=1pt, itemsep=0pt, partopsep=0pt, parsep=0pt]
\item{\em Linear Sum Assignment. } 
The objective term~\ref{eq:total_comm} and constraint~\ref{eq:balance_constraint} together form a linear sum assignment (LSA)~ \cite{lsa} problem, which can be solved using polynomial-time algorithms such as the Hungarian method~\cite{hungarian-algorithm} with time complexity $O(B^3)$. 
LSA methods have a faster and more predictable runtime than the ILP solver. It also
finds a strong initial solution that minimizes total communication  (equation~\ref{eq:total_comm}).

\item{\em Local search.} We use local search to iteratively refine the initial solution to improve load-balancing objectives. In each search step, we swap the GPU assignments of two image patches using a greedy steepest-ascent strategy. 
We iteratively perform greedy swapping until convergence or a time budget is reached.  To avoid getting stuck on plateaus caused by the non-smooth max terms, we replace them with $p$-norm and use the relaxed objective $\beta \cdot \|\texttt{send}\|_p + \gamma \cdot \|\texttt{recv}\|_p + \delta \cdot \|\texttt{comp}\|_p$, where $\|\texttt{send}\|_p = (\sum_{i=k}^{N}{\texttt{send}_k^p)})^{1/p}$ and so on. 
Appendix \ref{sec:appendix:stochastic-local-search} provides more details. 
\end{itemize}

We further fine-tune our LSA-based online optimization with two
important augmentations. First, instead of performing placement at the granularity of entire images, we 
divide each $H\times W$ image into $p^2$ patches of size $\frac{H}{p} \times \frac{W}{p}$ and determine the placement of individual image patches. We typically set $p$ to a small value $2$--$8$.  
Using finer-granularity allows for more flexible placement, which can result in reduced communication and improved load balance.  Ideally, we want to co-locate image rendering with most (if not all) of the points it accesses; doing so is rarely feasible for a whole image but more likely with a smaller image patch. Figure \ref{fig:design_pixelPatch} illustrates such an example. Also, training for \workloadabbrev uses small batch sizes~\cite{grendel}, $16$---$64$,  whole image assignment severely limited when scaling to many of GPUs. 

Second, we perform hierarchical images assignment similarly to hierarchical point placement. In particular, we first assign image patches across machines. We then assigns a machine's image patches among its local GPUs. The two levels of assignment use different objective coefficients $\alpha$, $\beta$, $\lambda$, and $\delta$. For intra-node assignment, we de-prioritize the total communication volume (i.e., by setting $\alpha$ to zero), as intra-node bandwidth is high.

\section{Implementation} 
\label{sec:implementation}

We implement \name with approximately 9K lines of Python, 3K of CUDA, and 1K of C++. 
Our system uses gsplat\cite{gsplat} as the rendering kernels backend, PyTorch\cite{pytorch} for both training and efficient tensor operators, and NCCL\cite{nccl} for communication. 
For bipartite graph partitioning, we use METIS \cite{metis}. 
Our CPU-based online image placement solver using linear sum assignment from SciPy\cite{SciPy}, and a custom parallel implementation of local search in C++ with OpenMP\cite{openMP}. 
We highlight two key components of our implementation; others are provided in Appendix \ref{sec:appendix:additional-implementations}.

\paragraph{Asynchronous online placement.} To hide the overhead of online assignment, we asynchronously compute image placement on the CPU for future batches while the GPU processes the current one. 
However, computing image placement requires the access pattern ($\mathcal{A}$) which is unavailable for batches whose training has not started.  
To approximate this statistic, we use a profiler to collect $\mathcal{A}$ for each image patch based on information from the previous epochs. Since points evolve gradually in training, these serve as reliable approximations for future epochs. 



\paragraph{Training dataset storage and loading.} To enable efficient image loading, we pre-decode the images and store them in CPU pinned memory, transferring them to the GPU on demand during each training iteration.
However, naively replicating the entire dataset in each machine would lead to CPU out-of-memory (OOM) errors, as the training dataset after decoding can be very large for high-resolution, large-scale scenes, often reaching hundreds of gigabytes. 
To address this, we assign each machine a fraction of the total images, so the training dataset is effectively stored in the aggregate CPU memory across all machines.  
During training, each GPU attempts to retrieve the required ground-truth images from its local subset. If an image is not found locally, it is fetched from another process that holds the corresponding subset via communication. 
The offline bipartite graph partitioning (\S\ref{subsubsec:pointplacement}) naturally produces a partitioning of images (in addition to points), which we can adopt here. 
Since online image assignment aim to co-locate camera views with their visible points, it is highly likely that a GPU is assigned images that already stored locally.

\section{Evaluation} 
\label{sec:eval} 

Our evaluations have the following highlighted results: 
\begin{compactitem}
    \item \name outperforms the state-of-the-art training system gsplat by 1.50–3.71$\times$ across 6 scenes and 3 \workloadabbrev methods. \name reduces communication volume by up to 91\%.
    
    \item \name scales \workloadabbrev training to 128 GPUs on MatrixCity Aerial, with 12.7$\times$ speedup from 8 to 128 GPUs. 
    
    \item \name enables large model training up to 500M points (29.5 billion parameters). To the best of our knowledge, this is the largest point cloud that has been successfully trained, yielding a new state-of-the-art reconstruction quality of 26.75 (test PSNR) on MatrixCity Aerial. 

\end{compactitem}

    

\subsection{Setting and Datasets} 

\paragraph{Experimental Setup.} 
We conduct our evaluation on a cluster of machines equipped with 4$\times$40GB NVIDIA A100 GPUs and an AMD EPYC 7763 CPU per machine. Intra-machine communication uses fully-connected third-generation NVLinks; while inter-machine communication relies on Ethernet with socket-based messaging (no RDMA), forming a hybrid networking topology. We utilize up to 32 machines, with point-to-point inter-machine bandwidth measured at 88 Gbps in one direction. The per-machine bandwidth is evenly divided among four GPUs. 

\paragraph{Datasets.} 
We evaluate \name using 6 public available datasets listed in Table~\ref{tab:scenes-main}
\footnote{``BigCity Street'' in this table refers to the bottom subregion within all streets of BigCity.} 
. 
These datasets are all large-scale, with up to 60,000 images per scene and resolutions as high as 6K. 
These scenes vary in scale, topology, and rendering resolution to represent a diverse set of training workloads.
The point cloud size is determined by the total amount of detail in each scene, which depends on both spatial area and image resolution. 
Increasing $P$ improves placement flexibility, but also increases optimization time (\S\ref{subsubsec:imageplacement}). Therefore, we set $P{=}2$ for Ithaca and BigCity Aerial due to their fast computation; and $P{=}4$ for the others, where slower GPU computation allows for better overlap. 
We adjust the batch size (BS) of training images and the number of GPUs used for distributed training based on the point cloud size and rendering resolution. 
We classify the datasets based on their capture setting. ``Rubble'', ``Sci-Art'', and ``BigCity Aerial'' are Aerial Datasets, with images collected from drones in a downward-facing view. ``Ithaca365'', ``Campus'', and ``BigCity Street'' are Street Datasets, captured from the ground. 
For each dataset, if the image poses or the initial point cloud is missing, we generate them using Colmap \cite{colmap}. 


\paragraph{Applications.} 
We evaluate \name on distributed training of three types of \workloadabbrev methods: 3D Gaussian Splatting (3DGS)\cite{3dgs}, 2D Gaussian Splatting (2DGS)\cite{2dgs}, and 3D Convex Splatting (3DCX)~\cite{3dcs}, which use 3D Gaussians, 2D Gaussians, and 3D Convex shapes as their point representations, respectively. 
3DGS is the most widely used, while 2DGS offers improved geometry reconstruction by using 2D Gaussians to better capture surface. 
2DGS and 3DGS share most attributes in their model representation. As for rendering pipeline, the main difference is that 2DGS has larger view-dependent states (20 vs. 11), and shows a higher rendering cost. 
In contrast, 3DCX differs significantly from both 3DGS and 2DGS. 
Instead of using position, scale, and rotation to define spatial coverage as in Gaussian-based methods, 3DCX defines coverage using the positions of its six vertices. Its rendering pipeline and hyperparameters are, accordingly, distinct. 
See Appendix \ref{sec:appendix:additional-background:difference-between-point-types} for more differences between these methods.

\begin{table}[h]
\centering
 \resizebox{\columnwidth}{!}{
\begin{tabular}{lrrrrrr}
    \toprule
    \textbf{Scene} & \textbf{\# Images} & \textbf{Res.}  & \textbf{PC Size}  & \textbf{P} & \textbf{\# GPU} & \textbf{BS} \\
    \midrule
    Rubble~\cite{meganerf} & 1600 & 4K & 40M & 4 &16 &  16 \\
    Ithaca365~\cite{ithaca365} & 8200 & 1K & 48M & 2 & 16 & 16 \\
    Campus~\cite{hierarchicalgaussians} & 20000 & 1K & 56M & 4 & 16 & 16 \\
    BigCity Street~\cite{matrixcity} & 34000 & 1K & 80M & 4 & 16 & 16 \\
    Sci-Art~\cite{UrbanScene3D, meganerf} & 3600 & 6K & 200M & 4 & 32 & 32 \\
    BigCity Aerial~\cite{matrixcity} & 60000 & 1080P & 100M & 2 & 32 & 64 \\
    \bottomrule
\end{tabular}
}
\caption{Scenes used in our evaluation. We list the number of images, which roughly reflects the scene area, and the resolution of training images. We also include the Point Cloud size (in Millions), image patch factor (P), GPU count, and batch size. Corresponding partitioning results are visualized in Figure \ref{fig:eval_pcd-partition-results} (Appendix \ref{sec:appendix:visualize-partitioning}). } 
\label{tab:scenes-main}
\end{table}



\paragraph{Baseline.} 
We choose gsplat \cite{gsplat} (v1.4.0) as our baseline, as it is the state-of-the-art \workloadabbrev training system. 
gsplat supports distributed training for 3DGS, but not for 2DGS or 3DCX. We extend gsplat's 2DGS and 3DCX rendering pipelines with distributed training support---a tedious process that requires identifying each kernel’s true parallel axis (per-point or per-pixel) and manually reconstructing a correct execution order in distribution. 
Unlike \name's locality-aware design, gsplat randomly distributes points and rendering tasks across GPUs. 
We encountered training bottlenecks at the dataloader, as gsplat streams images from disk and decodes them on the CPU by default. To address this, we augment the baseline with our efficient dataset loading implementation described in Section~\ref{sec:implementation}. Finally, both the baseline and \name use the same rendering kernels provided by gsplat. Both disable automatic Python garbage collection during training to prevent irregular disturbances, following the practice in~\cite{megascale}.

\subsection{Overall Performance}
\label{sec:eval:throughput}

\begin{figure*}[t]
    \centering
    \includegraphics[width=\linewidth]{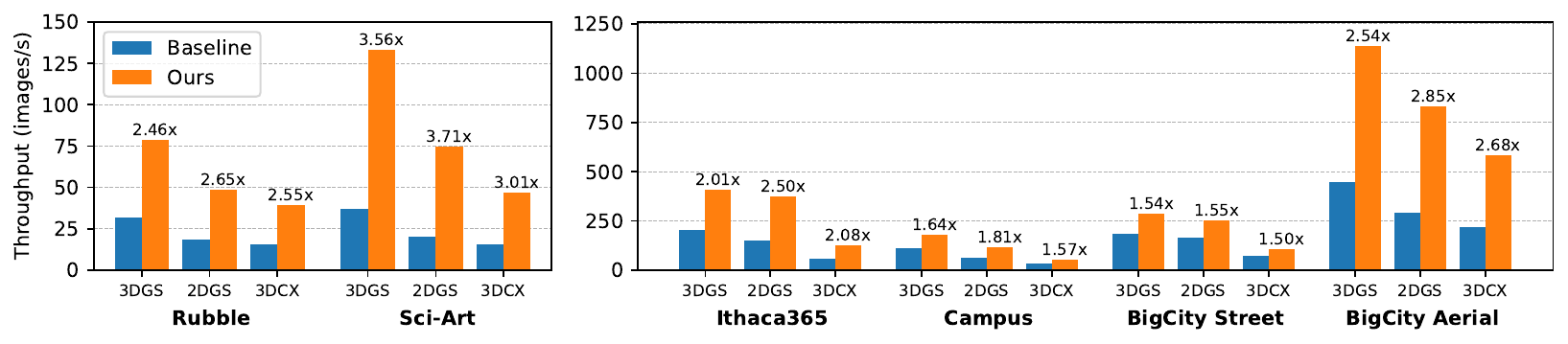}
    \caption{Training throughput (images/s) comparison across all scenes in Table~\ref{tab:scenes-main} using three types of \workloadabbrev methods: 3D Gaussian Splatting (3DGS), 2D Gaussian Splatting (2DGS), and 3D Convex Splatting (3DCX). The corresponding experimental settings are in Table~\ref{tab:scenes-main}. }
    \label{fig:throughput-all-scenes-all-primitives}
\end{figure*}

We first evaluate \name on a wide range of scene settings from Table~\ref{tab:scenes-main} and on all 3DGS, 2DGS and 3DCX methods, comparing against the baseline. 
Figure~\ref{fig:throughput-all-scenes-all-primitives} shows the average training throughput for each scene–method combination. 
For fairness, we use identical optimization hyperparameters for both the baseline and \name in each case. 
Averaged across scenes, \name outperforms the baseline by 2.30$\times$ (from 1.54 to 3.56$\times$) for 3DGS, 2.50$\times$ (from 1.55 to 3.71$\times$) for 2DGS and 2.23$\times$ (from 1.50 to 3.01$\times$) for 3DCX. These runtime statistics account for all aspect of training including overhead from online image placement and data loading if they are not fully overlapped. As such, these speedup numbers provide a holistic evaluation of our techniques. 

\name achieves better performance by reducing communication volume through improved data locality in distribution, and by mitigating both computation and communication load imbalance. 
\name yields greater speedup for 2DGS than for 3DGS. This is because 2DGS has higher communication overhead because its larger view-dependent states than 3DGS (20 vs 11 per point), making our communication-reduction techniques more effective. 
Although 3DCX communicates even more view-dependent states (29 per point), it yields the smallest average speedup. This is due to its distinct rendering pipeline, which shows much longer computation time (see their notably lower overall throughputs). As a result, the communication-to-computation ratio does not increase, and the benefits of communication reduction are limited.

We observe that the speedups achieved by \name vary across different scenes. 
In particular, \name’s locality-aware distribution performs better on Aerial Datasets, with an average speedup of 2.89$\times$, compared to 1.80$\times$ on Street Datasets. 
This disparity largely arises from their complex and irregular topology in Street Datasets, which make point cloud partitioning more challenging for \name (see Figure~\ref{fig:eval_pcd-partition-results} in Appendix~\ref{sec:appendix:visualize-partitioning}). Moreover, street-view frustums can span both nearby and distant content, compromising the data access locality that \name leverages. 
In contrast, Aerial Datasets typically exhibit more structured topologies---resembling a 2D plane (see Figure~\ref{fig:eval_pcd-partition-results})---and each view tends to cover only a small portion of the scene (e.g., less than 1\% in BigCity Aerial), enabling more effective partitioning and communication reduction. 
In addition, the absolute throughput varies significantly across scenes due to differences in data access volume, rendering resolution, and load imbalance. 
For example, scenes with higher rendering resolution exhibit lower training throughput. 

Behind the overall performance gains, we further observe minimal system overheads, as detailed in Appendix~\ref{sec:appendix:system-overheads}.




\paragraph{Communication Reduction.} 
To evaluate the data movement savings enabled by \name’s locality-aware distribution, we compare the communication volume during training between the baseline and \name. Table~\ref{tab:eval_comm-reduction} shows the average reduction in the number of points transferred across machines per iteration, achieved by \name. Across different scenes and applications, we observe communication reductions ranging from 53.8\% to 91.4\%. 

\begin{table}[h]
    \centering
    \resizebox{\columnwidth}{!}{
        \begin{tabular}{c c c c c c c}
            \toprule
            \multirow{2}{*}{\textbf{Scene}} & \multirow{2}{*}{\textbf{Rubble}} & \multirow{2}{*}{\textbf{Ithaca365}} & \multirow{2}{*}{\textbf{Campus}} & \textbf{BigCity} & \multirow{2}{*}{\textbf{Sci-Art}} & \textbf{BigCity} \\
            & & & & \textbf{Street} & & \textbf{Aerial} \\
            \midrule
            3DGS & 82.8\% & 89.2\% & 54.5\% & 57.7\% & 89.3\% & 89.9\% \\
            2DGS & 82.7\% & 84.0\% & 58.1\% & 55.5\% & 89.1\% & 89.9\% \\
            3DCX & 83.2\% & 80.8\% & 53.8\% & 54.8\% & 89.9\% & 91.4\% \\
            \bottomrule
        \end{tabular}
    }
    \caption{Inter-machine communication volume reduction achieved by \name across all scene-method combinations, relative to the baseline. These statistics correspond to the experiments presented in Figure~\ref{fig:throughput-all-scenes-all-primitives}.} 
    \label{tab:eval_comm-reduction}
\end{table}

We observe that the overall performance speedups reported in Figure~\ref{fig:throughput-all-scenes-all-primitives} positively correlate with the degree of communication reduction. 
Taking 3DGS as an example, we achieve the highest speedup on Sci-Art (3.56$\times$) with an 89.3\% reduction in communication volume, whereas on BigCity Street, the lowest speedup (1.54$\times$) corresponds to a 57.7\% reduction. 
Because communication is the bottleneck in baseline training (see Figure~\ref{fig:communication-challenge}), alleviating this core issue leads to significant speedup. 
Moreover, like the overall speedup differences, communication reduction also varies across scenes. 
Intuitively, communication reduction depends on the degree of data access locality during rendering. 
When each view’s frustum covers a compact region, it becomes easier to co-locate the necessary data on a single GPU, reducing communication. For example, in aerial scenes, all cameras point downward, and their captured regions are typically compact. 
In contrast, street-view scenes often involve frustums that extend beyond obstacles and capture distant content, resulting in a lower data locality, thereby causing more communication.

The percentage of communication reduction remains almost the same across point types. 
This is because, although different point types affect how scene details are represented, the global structure of the point cloud stays the same. 
Consequently, each view accesses points from similar regions regardless of point type, leading to comparable data access patterns and communication demands. 


\begin{figure}
    \centering
    \includegraphics[width=\linewidth]{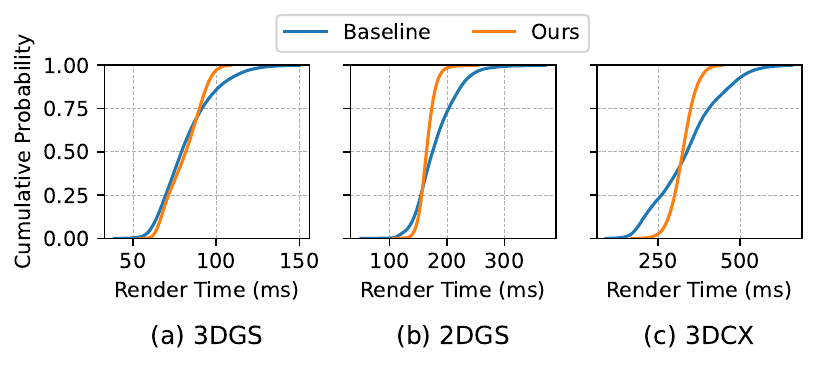} 
    \caption{Empirical CDF of rendering time for the Sci-Art dataset using 3DGS, 2DGS, and 3DCX. }
    \label{fig:eval_render_dist}
\end{figure}

\paragraph{Improvement on Rendering Load Balancing.}
To evaluate \name’s effectiveness in balancing rendering time across GPUs, we analyze the Sci-Art dataset, which has the highest image resolution among all scenes. 
Figure~\ref{fig:eval_render_dist} presents the empirical CDF of rendering times---measured as the sum of forward and backward passes with cuda events---collected from all GPUs throughout training, comparing 3DGS, 2DGS, and 3DCX. 
We observe that \name achieves lower variance in rendering time, as evidenced by the steep rise in its CDF curve---indicating that most GPUs spend a similar amount of time rendering. 
In contrast, the baseline curves exhibit more gradual slopes, reflecting greater variance of rendering time.

\subsection{Throughput Scalability}


\begin{figure}[t]
    \centering
    \begin{subfigure}{0.9\linewidth}
        \centering
        \includegraphics[width=\linewidth]{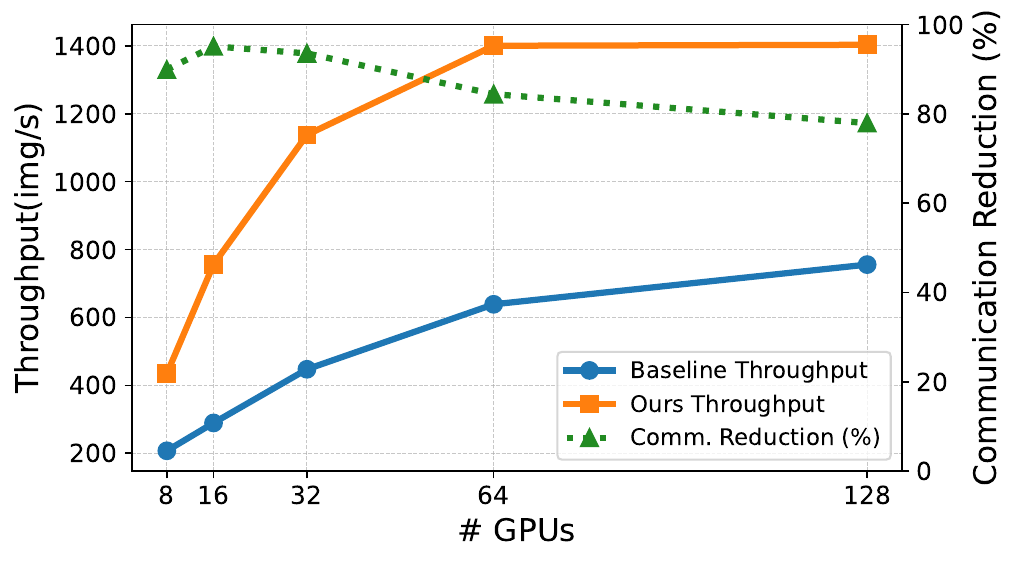}
        \caption{Strong scalability. We fix the global batch size to 64 images.}
        \label{fig:throughput_strong_scalability}
    \end{subfigure}
    \hfill
    \begin{subfigure}{0.9\linewidth}
        \centering
        \includegraphics[width=\linewidth]{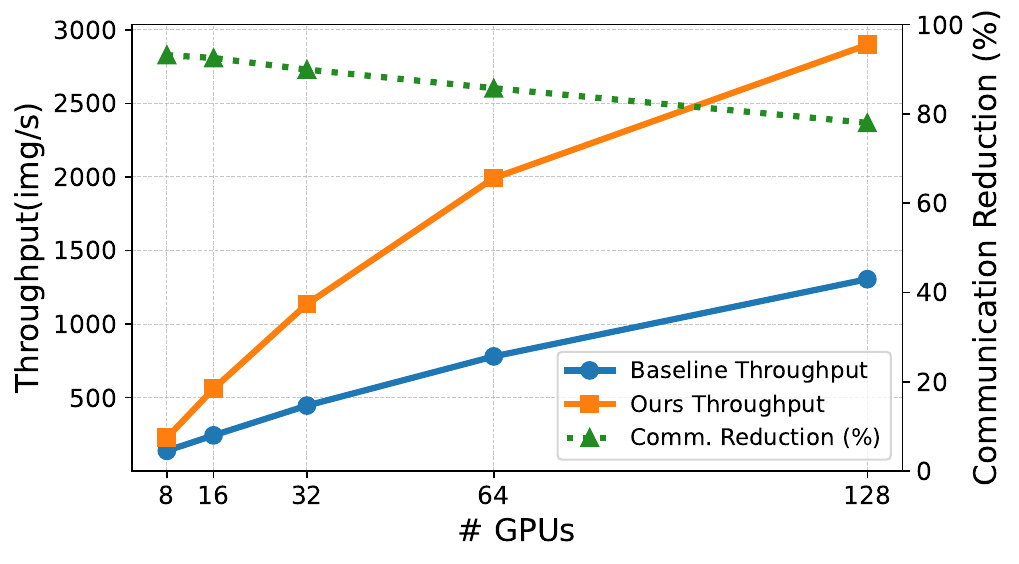}
        \caption{Weak scalability. We fix the per-GPU batch size to 2 images.}
        \label{fig:throughput_weak_scalability}
    \end{subfigure}
    \caption{
    Strong and weak scalability on Bigcity Aerial. We report training throughput for both the baseline and \name as the number of GPUs increases. Additionally, we show the percentage reduction in inter-machine communication achieved by \name. }
    \label{fig:throughput_scalability_combined}
\end{figure}

We evaluate the throughput scalability of \name on BigCity Aerial, using 3DGS as the rendering method. Experiments are conducted on 8, 16, 32, 64, and 128 GPUs. We assess both strong scalability (with a fixed global batch size of 64 images) and weak scalability (with a fixed per-GPU batch size of 2 images), comparing \name against the baseline. We also report the reduction in inter-machine communication volume achieved by \name relative to the baseline and analyze how this reduction trends across scales. 

\paragraph{Strong Scalability.} 
As shown in Figure~\ref{fig:throughput_strong_scalability}, \name consistently outperforms the baseline, achieving over 80\% reduction in communication volume along with multi-fold throughput gains. 
The throughput scaling of \name diminishes at higher GPU counts, as indicated by the decreasing slope of the curve. 
This aligns with \name’s communication reduction trend, which also declines with higher GPU counts. 
As the point cloud is split into more partitions (i.e., across more GPUs), the length of partition boundaries increases. 
This causes more views to span multiple partitions, which leads to more cross-GPU point access and thus higher communication overhead. 



\paragraph{Weak Scalability.} 
Under the weak scalability setting (Figure~\ref{fig:throughput_weak_scalability}), \name again consistently outperforms the baseline by a wide margin. 
The slope of the throughput curve also decreases slightly with scale. 
Throughput scales better under weak scalability, increasing from 226.9 img/s at 8 GPUs (batch size 16) to 2898.5 img/s at 128 GPUs (batch size 256). In contrast, under strong scalability (fixed batch size of 64), it only rises from 434.7 img/s to 1402.8 img/s. 


\begin{figure*}[t] 
    \centering 
    \includegraphics[width=\linewidth]{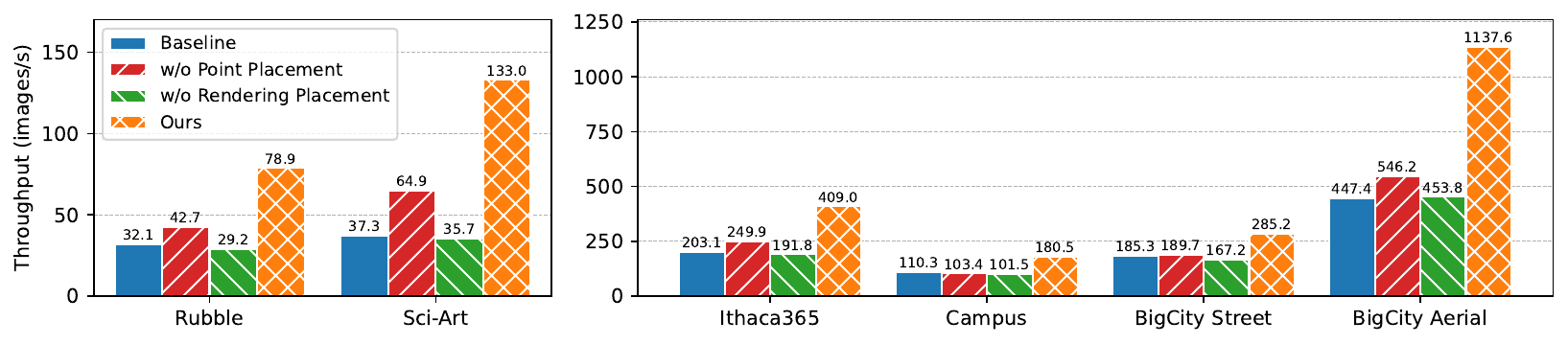} 
    \caption{Ablation study of locality-aware placement evaluated on 3DGS. ``w/o Point Placement'' randomly places points across GPUs (without designs in Section~\ref{subsubsec:pointplacement}), while still placing image patches with data-compute locality (with designs in Section~\ref{subsubsec:imageplacement}). ``w/o Rendering Placement'' does the reverse: points are placed with locality, but image patches are assigned randomly.  } 
    \label{fig:eval_ablation_placement} 
\end{figure*} 

\subsection{Ablation of Joint Point and Image Placement}

Our locality-aware distribution includes two components---point placement (Section~\ref{subsubsec:pointplacement}) and image rendering placement (Section~\ref{subsubsec:imageplacement})---both guided by data locality. We conduct an ablation study using the 3DGS method to evaluate their individual impact by measuring training throughput after disabling each component. 
Figure~\ref{fig:eval_ablation_placement} shows the results, indicating that locality-aware distribution is effective only when both \name's point and image placements are applied. 


Disabling only point assignment reduces \name’s training throughput by 33\%–52\% (BigCity Street and Sci-Art), as points are uniformly distributed across GPUs, causing each image to access similar amounts of data from all GPUs. 
This leaves little room for reducing data movement through strategic rendering image assignment. 
Even in this case, the online image placement still yields speedups over the baseline on the three high-resolution scenes---1.3$\times$ on Rubble and 1.74$\times$ on Sci-Art---but shows little to no improvement on lower-resolution datasets. This is because online image scheduling helps mitigate load imbalance when rendering high-resolution images that are more compute intensive. 

After disabling only the image assignment component, \name no longer outperforms the baseline. This is expected---randomly assigning images fails to exploit data locality, even if points have been stored with spatial locality.  
As a result, communication is not reduced and may even become more imbalanced due to uneven point distribution. 
Due to page limit, we report the ablation study of three additional techniques---hierarchical assignment, load balancing constraints, and patch factor---in Appendix~\ref{sec:appendix:additional-ablation}. 
All three components contribute to performance; and even when one is removed, \name still has significant speedup. 








\subsection{Training Larger Models and better PSNR} 

\begin{figure}[t]
    \centering
    \includegraphics[width=0.84\linewidth]{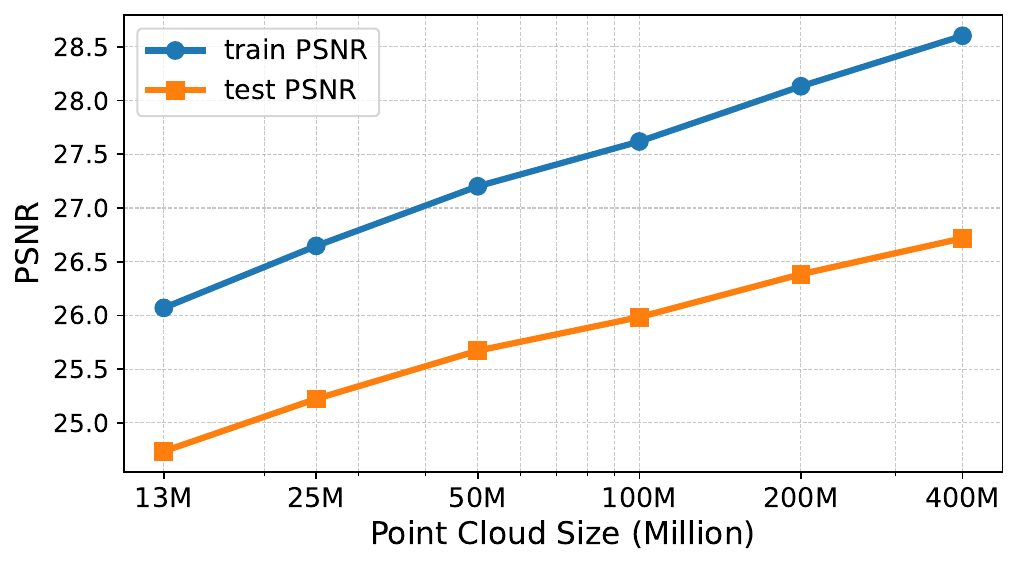}
    \caption{Reconstruction quality (PSNR) vs model size with 3DGS on the BigCity Aerial scene. All models are trained on 8 machines for 1.5 million steps. } 
    \label{fig:psnr_scalability_BigCity_aerial}
\end{figure}

\name is the first system to successfully train on the full MatrixCity BigCity dataset~\cite{matrixcity} using \workloadabbrev, overcoming its extreme memory and computational demands. We use PSNR (Peak Signal-to-Noise Ratio) to evaluate reconstruction quality; it is a standard image reconstruction metric~\cite{nerf,3dgs,3dgssurvey}, where higher values indicate better fidelity. 
For the aerial view (60K images) in BigCity, \name trains a 400M-point model on 8$\times$4 A100 GPUs, achieving a PSNR of 26.75 on test set and a PSNR of 28.60 on training set. 
For the street view (260K images), \name trains a 500M-point model on 16$\times$4 A100 GPUs, reaching a test PSNR of 21.29 and a train PSNR of 25.38. These models contain 23.6B and 29.5B parameters respectively (based on 3DGS with 59 attributes per point), far exceeding single-machine GPU memory limits. 
This large point cloud size is essential for achieving the reconstruction quality shown in Figure~\ref{fig:psnr_scalability_BigCity_aerial}. 
Reconstruction quality (PSNR) improves with model size, as finer details are captured. 
See Appendix~\ref{sec:appendix:training-matrixcity} for additional training details.

\subsection{Dynamic Scene Reconstruction for 3D Video} 

\workloadabbrev can be generalized to reconstruct 3D videos where every frame in the video is a static 3D scene viewable from arbitrary viewpoints.  The required training dataset consists of multi-view videos where each frame is annotated not only with a camera pose but also with a timestamp.   We build a popular video reconstruction algorithm, 4D Gaussian Splatting (4DGS)~\cite{yang4dgs}, and evaluate its performance. 
4DGS extends the point type to a 4D Gaussian by adding a temporal dimension. Each point possesses attributes that define its lifespan and movement, allowing the model to represent scene evolution over time.  Appendix~\ref{sec:appendix:additional-examples} shows our 4DGS implementation~\ref{sec:appendix:additional-examples}.
The key distinction lies in how 4DGS implements {\sf pts\_culling}. In static reconstruction, points are culled solely if they fall outside the view frustum; in dynamic reconstruction, points are culled if they are either outside the frustum or not present at the requested rendering timestamp. 

\begin{figure}[t]
    \centering
    \includegraphics[width=0.84\linewidth]{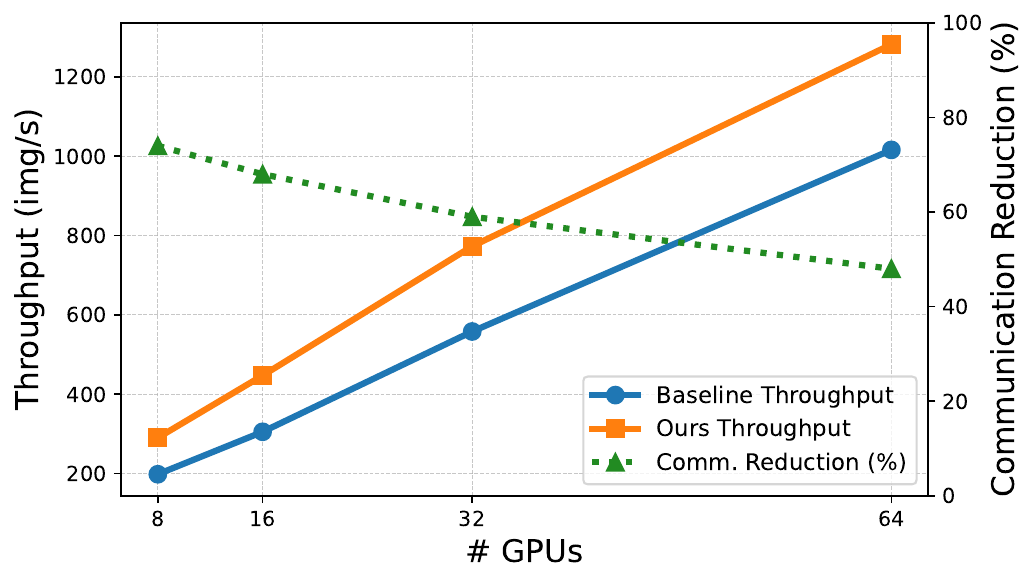}
    \caption{Weak scalability of \name on 3D video reconstruction, with per-GPU batch size fixed at 4.}
    \label{fig:weak_scalability_4dgs}
\end{figure}

We evaluate 4DGS using the Corgi dataset \cite{longvolcap}. 
It contains 24 one-minute videos of the same dynamic scene captured from distinct camera positions, which are downsampled to 30 FPS at 1K resolution for training. 
Our evaluation uses a model with 29M-points, which cannot be trained on a single GPU due to out-of-memory (OOM). \name enables scalable distributed training. As shown in Figure~\ref{fig:weak_scalability_4dgs}, the number of GPUs is increased from 8 to 64, \name improves training throughput by 4.42$\times$. 
Furthermore, compared to random distribution, \name's locality optimization significantly improves performance: at 64 GPUs, it yields throughput improvement of 26\% while reduces inter-machine communication by 48\%.

\section{Related Works} 
\label{sec:related}

\paragraph{Distributed 3D Gaussian Splatting Training. } 
Several approaches~\cite{grendel,dogs,retinags} have explored multi-GPU training for 3DGS, but their designs are specific to 3DGS and do not provide abstractions that generalize to other \workloadabbrev. 
Grendel~\cite{grendel} and gsplat~\cite{gsplat} randomly distribute computation in both point and image dimensions; 
\name builds on them with locality-aware optimization and a general abstraction. 
Moreover, DoGS~\cite{dogs} relies on asynchronous updates that may affect convergence; RetinaGS~\cite{retinags} incurs redundant compute and memory through replicated rendering; and CLM~\cite{clmgs} scales training through CPU offloading at the cost of efficiency. 
None of these systems report results at the scene scale achieved in our work.

\paragraph{Locality-aware distributed computation and scheduling.} 
Existing systems exploit locality in both classic big-data computation and distributed ML training. 
MapReduce~\cite{mapreduce} on GFS/HDFS~\cite{gfs,hdfs} assigns map tasks to machines holding their input blocks, while Purlieus~\cite{Purlieus} further optimizes reduce-task placement.
Other systems improve locality at the cluster scheduler: Delay Scheduling \cite{delayscheduling} temporarily skips jobs without local data; Quincy \cite{quincy} and Firmament \cite{firmament} formulate locality-aware scheduling as a minimum-cost flow problem for holistic optimization. 
These approaches place computation on a fixed data layout, while complementary work modifies data placement itself: Scarlett~\cite{scarlett} replicates hot HDFS blocks. 
\name follows the same high-level intuition of moving computation closer to data. 
However, these systems only leverage locality at coarse granularities (jobs, tasks, or storage blocks). 
Ray~\cite{Ray} similarly applies locality heuristics only at abstract objects and tasks levels. 
In contrast, \name optimizes fine-grained locality induced by the geometric visibility between points and camera views in \workloadabbrev. 
Moreover, whereas these systems often face unpredictable access patterns, \workloadabbrev exposes fully predictable access and locality: training views are predetermined, and lightweight frustum culling precisely identifies the points each view accesses. 
In distributed ML workloads such as GNNs and DLRMs, the training set is also fixed, making data access in principle also predictable. 
However, their data accesses follow a highly skewed, power-law distribution \cite{PowerGraph,UGache,distGNN,BGL,rethinkinggraphdataplacement,Bagpipe,Herald}, so systems primarily optimize locality by placing and caching hot items. 
GNN systems such as Dorylus \cite{dorylus}, BGL \cite{BGL}, DistGNN \cite{distGNN} and others \cite{PaGraph,rethinkinggraphdataplacement} exploit graph locality by partitioning nodes/edges and replicating or caching frequently accessed features so that most neighbor lookups remain local. 
P3 \cite{p3distgnn} reduces communication by avoiding feature movement rather than explicitly placing hotspots, yet its execution remains shaped by the same power-law neighborhood structure. 
For DLRM, recent systems mitigate massive, skewed embedding tables by keeping hot IDs near trainers. BagPipe~\cite{Bagpipe} uses lookahead caching, Herald~\cite{Herald} applies skew-aware scheduling, and HypeReca~\cite{HypeReca} replicates a small GPU hot set.
UGache~\cite{UGache} unifies GNN and DLRM by modeling their power-law embedding accesses as a skew-driven locality problem with optimized placement.
In contrast, \workloadabbrev is fundamentally different: point–view accesses are predictable but not skewed. Locality therefore cannot be captured by hotspot optimizations and must instead be exploited over the full, fine-grained geometric visibility structure between points and camera views.


\paragraph{Distributed Neural Network Training. } 
Many distributed training systems target neural networks, including data, tensor, pipeline parallelism~\cite{pytorch_dp, megatronlm, megatronlm2, gpipe, pipedream}, and FSDP-style methods~\cite{FSDP, deepspeed}. Some further support hybrid parallelism or automated parallelization~\cite{tofu, alpa, gspmd}.
These systems target workloads dominated by regular tensor operations (e.g., GEMM) and dense, regular communication (e.g., AllReduce). 
In contrast, rendering in \workloadabbrev training is irregular, with sparse, highly non-uniform data access that induces dynamic, irregular all-to-all communication.

\paragraph{Accelerating \workloadabbrev.} 
Recent work accelerates 3DGS via specialized hardware or software.
GScore \cite{gscore}, ACR \cite{ACR}, GauSPU \cite{GauSPU}, and MetaSapiens \cite{metasapiens} introduce custom hardware accelerators, while FlashGS \cite{flashgs}, StopthePop \cite{stopthepop}, BalancedGS \cite{balanced3dgs} and TamingGS \cite{taming3dgs} speed up rendering in software. 
These methods accelerate computation itself, whereas \name preserves the original kernels and instead reduces communication and load imbalance in distributed training.
Our approach is complementary, as we can directly leverage their accelerators; and faster computation (e.g., 15.6$\times$ in GScore) further amplifies communication as the dominant bottleneck, increasing the benefit of our communication-reduction techniques. See Appendix \ref{sec:appendix:additional-related-works} for more related works.

\section{Conclusion}
\label{sec:concl}
\name addresses key limitations in existing distributed \workloadabbrev training by introducing a general interface to abstract different \workloadabbrev algorithms and a novel locality-aware distribution strategy. By carefully optimizing both point placement and image assignment through a combination of offline and online techniques, \name significantly reduces communication overhead. \name supports multiple \workloadabbrev algorithms, scales to hundreds of GPUs, and enables training of unprecedentedly large models—pushing the boundaries of 3D scene reconstruction.

\bibliographystyle{ACM-Reference-Format}
\bibliography{08_grendelxl-bib}


\appendix
\newpage\clearpage

\section{Additional Background} 

\subsection{Differences in Rendering Across Point Types} 
\label{sec:appendix:additional-background:difference-between-point-types} 

\begin{table}[h]
    \centering

    \begin{subtable}[t]{\columnwidth}
        \centering
        \resizebox{\columnwidth}{!}{
            \begin{tabular}{c c r l}
                \toprule
                \textbf{State} & \textbf{\# Elem.} & \textbf{Size} & \textbf{Description} \\
                \midrule
                means2d & 2 & 8 B & 2D projected position of the Gaussian in camera space \\
                conics & 3 & 12 B & Encodes the 2D elliptical footprint (conic) of the Gaussian \\
                 &  &  & in screen space \\
                opacities & 1 & 4 B & Alpha transparency used in blending during rendering \\
                colors & 3 & 12 B & RGB color used to shade the Gaussian \\
                radii & 1 & 4 B & Scalar indicating size in screen space, used for LOD and culling \\
                depths & 1 & 4 B & Depth from the camera; used for depth sorting and visibility \\
                \midrule
                Total & 11 & 44 B & \\
                \bottomrule
            \end{tabular}
        }
        \caption{3D Gaussian Splatting (3DGS).}
        \label{tab:states-3dgs}
    \end{subtable}

    \vspace{1em}

    \begin{subtable}[t]{\columnwidth}
        \centering
        \resizebox{\columnwidth}{!}{
            \begin{tabular}{c c r l}
                \toprule
                \textbf{State} & \textbf{\# Elem.} & \textbf{Size} & \textbf{Description} \\
                \midrule
                means2d       & 2  & 8 B  & 2D projected position of the Gaussian in pixel space \\
                ray\_transforms & 9  & 36 B & 3×3 affine matrix mapping pixel‐space UV to \\
                 &  &  &  splat coordinates (row‐major KWH) \\
                opacities     & 1  & 4 B  & Alpha transparency value for blending the Gaussian \\
                colors        & 3  & 12 B & RGB color used to shade the Gaussian \\
                radii         & 1  & 4 B  & Maximum projected radius in pixels for culling/LOD \\
                depths        & 1  & 4 B  & Z‐depth of the Gaussian center for depth sorting \\
                normals       & 3  & 12 B & Surface normal in camera space for lighting/back‐face culling \\
                \midrule
                Total         & 20 & 80 B &  \\
                \bottomrule
            \end{tabular}
        }
        \caption{2D Gaussian Splatting (2DGS).}
        \label{tab:states-2dgs}
    \end{subtable}

    \vspace{1em}

    \begin{subtable}[t]{\columnwidth}
        \centering
        \resizebox{\columnwidth}{!}{
            \begin{tabular}{c c r l}
                \toprule
                \textbf{State} & \textbf{\# Elem.} & \textbf{Size} & \textbf{Description} \\
                \midrule
                means2d            & 2  & 8 B   & 2D projected center of the convex in pixel space \\
                normals            & 12 & 48 B  & 2D outward normals for each of the 6 projected hull vertices \\
                offsets            & 6  & 24 B  & Line‐offsets for each hull edge in the 2D indicator function \\
                opacities          & 1  & 4 B   & Alpha transparency for blending the convex splat \\
                colors             & 3  & 12 B  & RGB color coefficients for shading \\
                radii              & 1  & 4 B   & Projected radius (pixels) for culling \\
                depths             & 1  & 4 B   & Z‐depth of the convex center for depth sorting \\
                delta              & 1  & 4 B   & Smoothness parameter controlling edge curvature $(\delta)$ \\
                sigma              & 1  & 4 B   & Sharpness parameter controlling boundary transition $(\sigma)$ \\
                points view        & 1  & 4 B   & Number of vertices in the 2D convex hull ($\le 6$) \\
                \midrule
                Total              & 29 & 116 B & \\
                \bottomrule
            \end{tabular}
        }
        \caption{3D Convex Splatting (3DCX). 
        } 
        \label{tab:states-3dcx}
    \end{subtable}
    \caption{Comparison of view-dependent states across 3DGS, 2DGS, and 3DCX in rendering. These states are exchanged across GPUs in distributed training. Each point incurs a state size of 40~B, 80~B, and 116~B, respectively. For each state, we report the number of elements and their size in bytes (B), along with a brief description. } 
    \label{tab:view-dependent-states}
\end{table}

Different point types are defined by different attributes. 
For example, to define position, a 3D Gaussian uses \texttt{xyz} (3 $\times$ 32-bit values) for its center, while a 3D Convex uses 6 vertices, each with \texttt{xyz}, totaling 18 $\times$ 32-bit values.
Moreover, while the rendering operations can always be grouped into the APIs shown in Figure~\ref{fig:sys-api}, their specifics vary across methods. 
In 3DGS, projecting a 3D Gaussian is just a perspective projection. In contrast, 3DCX needs an extra step: after projecting its 6 vertices, it runs a Graham-scan to create a 2D space convex hull. 
The projection itself can also differ: 3DGS uses a simple affine approximation of perspective projection, while 2DGS models it more accurately, at the expense of increased computation and memory usage. 


Rendering operations yield varying view-dependent states in different methods. 
The three methods in our evaluation, 3DGS, 2DGS, and 3DCX incur view-dependent state sizes of 11, 20, and 29 elements per point, respectively. We break down the components of these view-dependent  states in Tables~\ref{tab:states-3dgs}, \ref{tab:states-2dgs}, and \ref{tab:states-3dcx}. 
Since our techniques primarily target communication reduction---which occurs during the exchange of view-dependent states---the effectiveness of reduction depends on the size of these states. Larger view-dependent states result in higher communication overhead, offering more opportunities for reduction.

\section{Additional Algorithm Implementations} 
\label{sec:appendix:additional-examples} 

In this section, we further describe how 3DCX and 4DGS are implemented using \name’s API, and how they are different from 3DGS. 

\begin{figure}[h] 
    \centering
    \includegraphics[width=\linewidth]{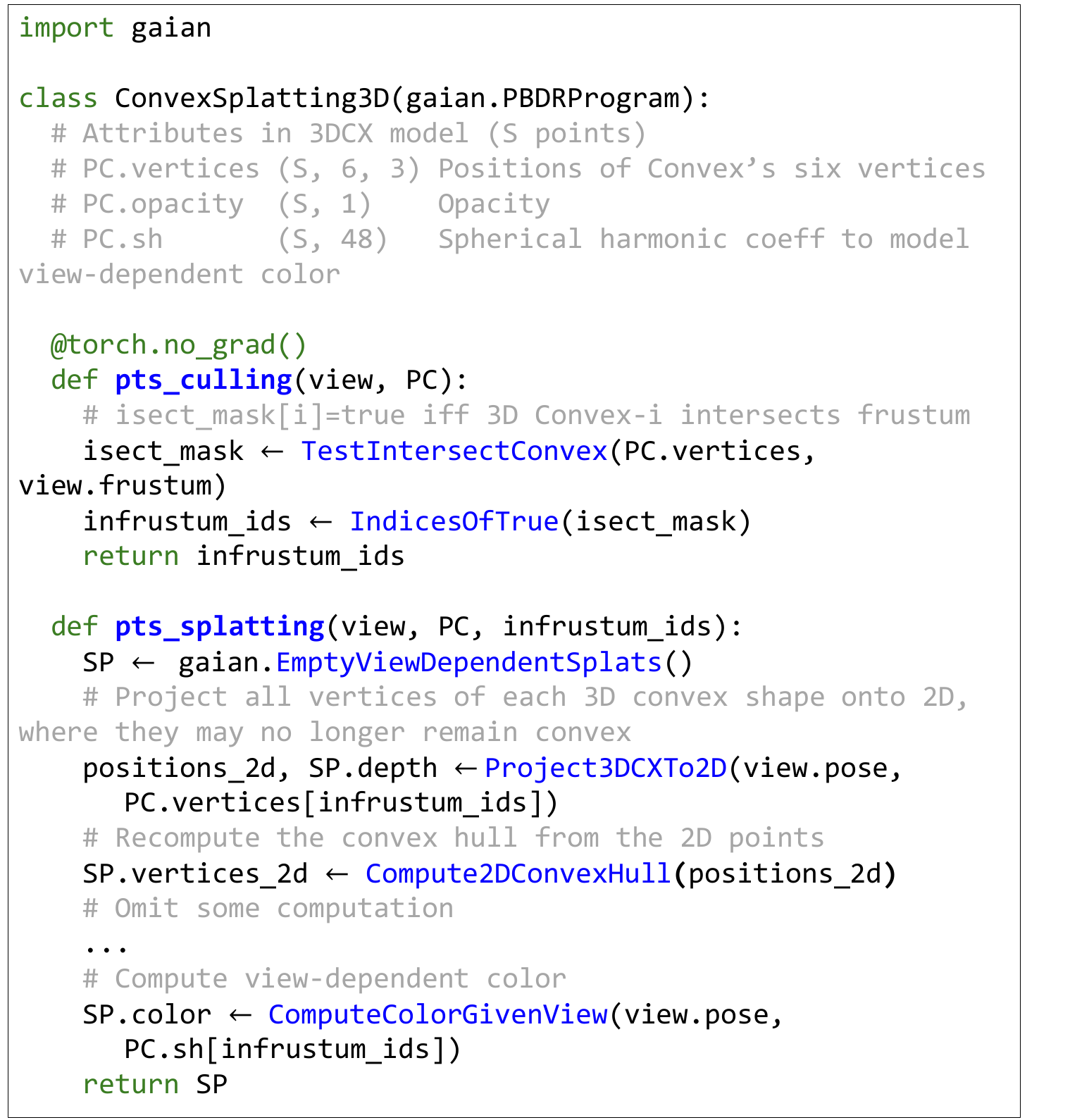} 
    \caption{Implement 3DCX with \name API.  } 
    \label{fig:3dcx-implement-api}
\end{figure}
\paragraph{3DCX. } 
As shown in Figure~\ref{fig:3dcx-implement-api}, each point in 3DCX is represented as a 3D convex polyhedron defined by six vertices, without the scale or rotation parameters used in 3DGS. So, in \textcolor{blue}{{\sf pts\_culling}}, this leads to a different frustum-intersection test.  
In the subsequent \textcolor{blue}{{\sf pts\_splatting}} stage, \textcolor{blue}{{\sf pts\_splatting}} projects all six vertices of in-frustum 3D convex onto the image plane and then recomputes a 2D convex hull as the view-dependent states. 
Moreover, 3DCX shares the opacity and spherical harmonics coefficients used in 3DGS, and these attributes undergo the same transformations during \textcolor{blue}{{\sf pts\_splatting}}. 
Finally, 3DCX produces view-dependent splat state ({\sf SP}) that differ from those of 3DGS; thus they require different \textcolor{blue}{{\sf image\_render}} pipelines. 
For simplicity, we omit two secondary attributes here used for modeling smoothness.

\begin{figure}[h] 
    \centering
    \includegraphics[width=\linewidth]{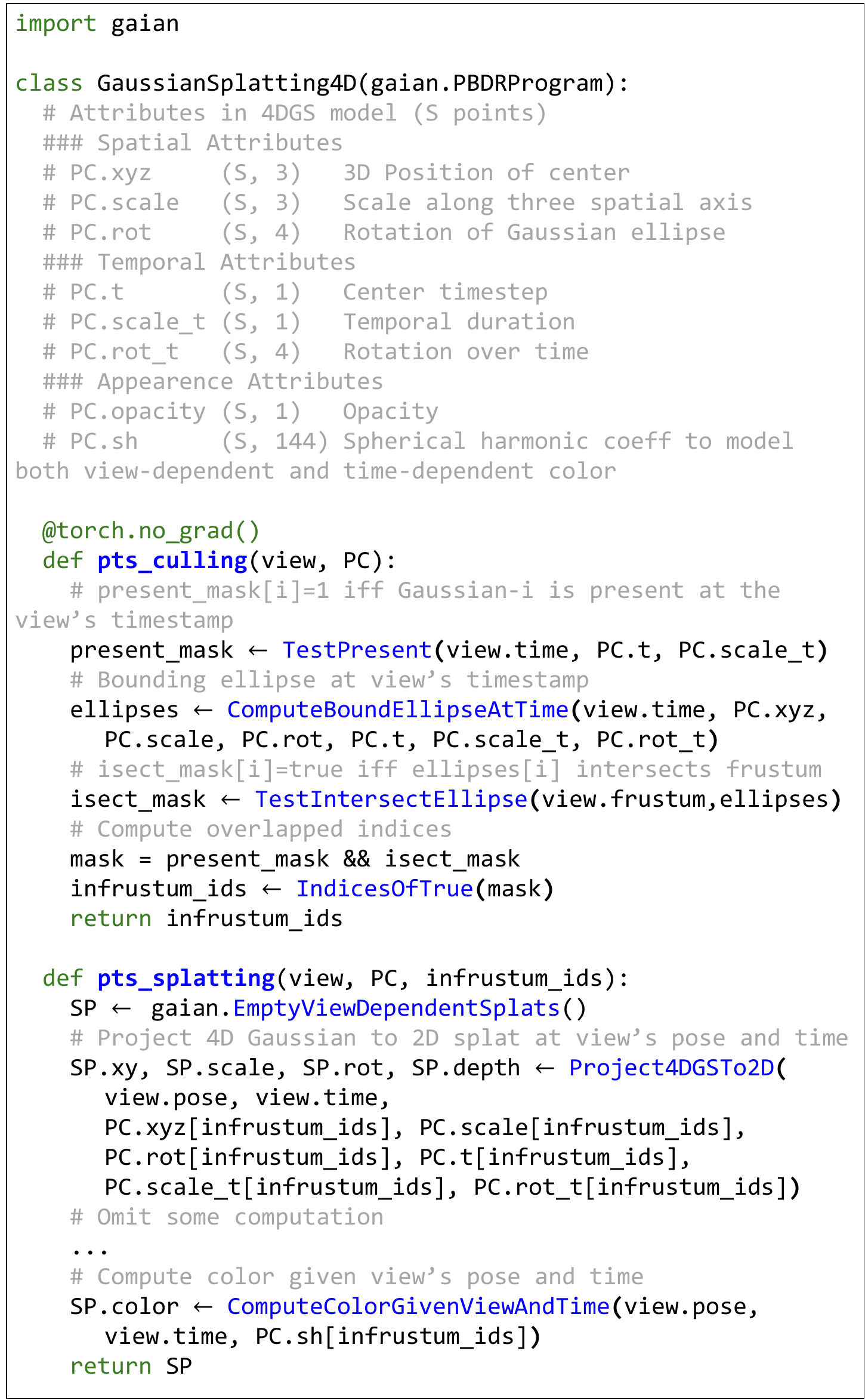} 
    \caption{Implement 4DGS with \name API. } 
    \label{fig:4dgs-implement-api}
\end{figure}
\paragraph{4DGS. } 
Different from 3DGS and 3DCX, which are used to reconstruct a static scene, 4DGS reconstructs a 3D video where the scene evolves over time. In this problem setting, each camera view includes not only a pose ({\sf view.pose}), but it also has a timestamp ({\sf view.time}) indicating when the view was captured. 
To model the dynamics, 4DGS replaces the 3D Gaussian used in 3DGS with a 4D Gaussian that incorporates time as an additional dimension. 
As shown in Figure~\ref{fig:4dgs-implement-api}, it introduces {\sf PC.t} and {\sf PC.scale\_t} to define each point’s presence interval over time; {\sf PC.rot\_t} governs shape deformation across time; and its spherical harmonics coefficients expand substantially (144 vs.\ 48 parameters in 3DGS) to additionally capture time-dependent color changes. 
Despite being a different reconstruction problem, 4DGS can still expose its data-access patterns through \textcolor{blue}{{\sf pts\_culling}}. Specifically, points are culled not only when their bounding ellipse does not intersect the camera frustum ({\sf isect\_mask}), but also when they are not present at the queried view’s timestamp ({\sf present\_mask}). 
The \textcolor{blue}{{\sf pts\_splatting}} stage follows the same high-level structure as in 3DGS: projecting points and evaluating spherical harmonics–based color. The key difference is that 4DGS performs projection from 4D to 2D and conditions the color computation additionally on the queried time.  
The resulting view-dependent splat state match those in 3DGS, allowing it to reuse 3DGS’ \textcolor{blue}{{\sf image\_render}} implementation.

\section{Additional Design Details} 




\subsection{Local Search} 
\label{sec:appendix:stochastic-local-search}

\paragraph{Local Search and Objective Relaxation.} 
We use the Linear Sum Assignment (LSA) solution as an initial solution that can minimize total communication. Starting from this, we apply a local search to further reduce load imbalance without significantly increasing communication cost.  

At each step of the search process, we select a pair of images and swap their currently assigned GPUs. We adopt a simple steepest-ascent strategy, where each step chooses the swap that yields the greatest improvement in the below objective function. While more advanced algorithms such as Simulated Annealing, Tabu Search, or Evolutionary Algorithms are also applicable, we find the hill climbing approach to be sufficient for our setting. 
We iteratively perform this swapping step until a satisfactory solution is found or the online placement time budget is reached. 

Max terms in the original objective function \ref{eq:optimization-formula} introduce non-smoothness and discrete transitions, leading to a rugged search landscape with plateaus and sharp jumps. This makes local search methods prone to getting stuck in suboptimal regions. 
Therefore, we smooth the max terms using the $p$-norm $\|\cdot\|_p$. When $p = \infty$, it recovers the maximum, i.e., $\|x\|_\infty = \max_i |x_i|$; when $p = 1$, it becomes the sum, i.e., $\|x\|_1 = \sum_i |x_i|$. Choosing $p \in (1, \infty)$ balances total and maximum values, allowing us to unify total communication and peak send/receive terms in the objective. Finally, we simplify the objectives function: 
\[
\begin{aligned} 
\text{minimize} \quad 
& \beta \cdot \| \texttt{send} \|_p 
+ \gamma \cdot \| \texttt{recv} \|_p 
+ \delta \cdot \| \texttt{comp} \|_p.
\end{aligned}
\]


This objective is not only smoother, making optimization easier; but it also aligns well with our swapping-based search strategy by helping accelerate online image placement. The objective can be quickly updated after each swapping step, and these swapping trials can be parallelizable (i.e. OpenMP multi-threads), both contributing to a faster search. Moreover, early stopping allows us to easily control the overall search time. 
Together, these features enable low-latency online placement. 

To set the objective coefficients for local search, \name’s Profiler measures average communication \(T_{\texttt{comm}}\) and average rendering computation times \(T_{\texttt{comp}}\) from previous epochs. 
It sets $\gamma = \frac{T_{\texttt{comp}}}{T_{\texttt{comm}} + T_{\texttt{comp}}}$ and $\alpha + \beta + \gamma = \frac{T_{\texttt{comm}}}{T_{\texttt{comm}} + T_{\texttt{comp}}}$. 
This prioritizes communication related objectives when communication dominates, and prioritizes computation related ones when computation dominates. 
Since local search relaxes the three communication terms into two $p$-norms, we only split the weight between $\beta$ and $\gamma$. 
We empirically choose $p$ based on the severity of load imbalance---scenes with more irregular topology use a larger $p$ to better reflect imbalance. The choice of $p$ is used to capture the effect of $\alpha$. 
Given the current maximum send/receive volumes, the weights are: \(\beta = \frac{\texttt{max\_recv}}{\texttt{max\_recv} + \texttt{max\_send}} \cdot \frac{T{\texttt{comm}}}{T{\texttt{comm}} + T{\texttt{comp}}}\) and \(\gamma = \frac{\texttt{max\_send}}{\texttt{max\_recv} + \texttt{max\_send}} \cdot \frac{T{\texttt{comm}}}{T{\texttt{comm}} + T{\texttt{comp}}}\).

\section{Additional Implementations} 
\label{sec:appendix:additional-implementations} 

\subsection{Customized Frustum Culling Kernel} 


In the original implementation of \workloadabbrev training~\cite{gsplat, 3dgs}, the Frustum Culling kernel has a complexity of $\mathcal{O}(B \times S)$, where $S$ is the total number of points. This is because each point must undergo a coordinate transformation for every view to determine whether it lies within the frustum. 
We optimize this step by leveraging Z-order linearization. Specifically, we compute an axis-aligned bounding box (AABB) for each contiguous block of points (Point Group). For each view, we first check whether any of the eight corners of the bounding box falls within the view frustum. If none do, we cull the entire Point Group, avoiding per-point checks. This reduces the number of culling operations from the number of points in a Point Group (e.g., 4096) to just 8 corner checks. 
Because each Point Group is spatially localized, this filtering is highly effective. As a result, the time complexity becomes proportional to the total number of in-frustum Point Groups across all camera views. 
Since in-frustum points constitute only a small fraction of the entire point cloud, this leads to highly efficient frustum culling. 

\subsection{Kernels for Rendering Image Patches}

\name leverages the rendering kernels from gsplat~\cite{gsplat}. However, the original gsplat kernels are designed to render entire images, rather than specific image patches within an image. We restrict rendering to specified regions by modifying CUDA kernels throughout the rendering pipeline for all 3DGS, 2DGS and 3DCX which are used in our evaluation.

\section{Additional Experiments} 






\subsection{Training MatrixCity BigCity} 
\label{sec:appendix:training-matrixcity} 

\paragraph{Training 400 Million points model for aerial view.} 
We train a 400 million-point model with \name using 32 A100 GPUs on the MatrixCity aerial dataset, which provides 60,000 images as the training set. We use the dataset’s official test set for evaluation and achieve a PSNR of 26.75 on the test set and 28.60 on the training set. All of \name’s techniques are enabled. Due to computational cost, training on Aerial dataset is performed for a total of 150,000 image steps. This number of steps is sufficient for smaller models (fewer than 50 million points) to converge in terms of PSNR. To ensure a fair comparison in the reconstruction quality scaling experiment with respect to model size (Figure~\ref{fig:psnr_scalability_BigCity_aerial}), we use the same number of training steps for all model sizes. 
However, we observe that the 400 million-point model does not fully converge within 150,000 steps, and will benefit from extended training. 
We use the default hyperparameters from gsplat, along with the selective Adam optimizer specifically designed for \workloadabbrev in gsplat. 

\paragraph{Training a 500 Million-Point Model for the Street View.}
We train a 500 million-point model with \name using 64 A100 GPUs on the MatrixCity Street dataset, which provides 260,000 images as the training set. We use the dataset’s official test set for evaluation and achieve a PSNR of 21.29 on the test set and 25.38 on the training set. All of \name’s techniques are enabled.
Training is conducted for a total of 5,500,000 image steps. We adopt the selective Adam optimizer. 
The hyperparameter tuning is more complicated, as street-view scenes are generally more difficult to reconstruct. We found the following settings to be important: we set the far plane to $\frac{1}{16}$ of the scene scale to exclude distant content, and apply a radius clip of 3 pixels to discard projected points that are too small. Training was stopped after 5,500,000 steps because of computational costs, although convergence was not yet observed. 
The gap between training and test PSNR is notably larger in the street view than in the aerial view. We suspect two contributing factors. First, the training and test sets are sampled from different camera trajectories (extrapolation), rather than from the same trajectory (interpolation)~\cite{matrixcity}, making generalization more challenging. Second, the MatrixCity BigCity Street dataset employs auto-exposure during image capture, resulting in varying exposure conditions that require additional vision algorithm to properly handle. 



\subsection{Additional Ablation Study} 
\label{sec:appendix:additional-ablation} 
We conduct additional ablation study to evaluate the contribution of three designs of our system across all scenes, using 3DGS as the rendering method. The three ablation variants are defined as follows: 

\paragraph{w/o Hier:} Disable hierarchical assignment of points and images in \name, which is first assigned between machines, then between GPUs within each machine. Instead, the point cloud is directly partitioned into the total number of GPU parts, with each part assigned to a single GPU. 

\paragraph{w/o Load Bal:} Disable the load balance of the point assignment and the image assignment in \name. 

\paragraph{w/o Patch:} We disable the assignment of patches smaller than a full image during training, as described in Section~\ref{subsubsec:imageplacement}. The offline partitioning and online assignment algorithms remain the same.


\begin{table}[h]
\centering
\resizebox{\columnwidth}{!}{
    \begin{tabular}{lrrrrr}
    \toprule
    \textbf{Scene} & \textbf{Baseline} & \textbf{Ours} & \textbf{w/o Hier.} & \textbf{w/o Load Bal.} & \textbf{w/o Patch} \\
    \midrule
    Rubble & 32.13 & \underline{78.90} & 50.57 & 72.98 & 45.29 \\
    Ithaca365 & 203.09 & \underline{466.48} & 391.83 & 462.92 & 420.78 \\
    Campus & 110.34 & \underline{180.54} & 125.26 & 168.09 & 135.92 \\
    BigCity Street & 185.28 & \underline{285.18} & 236.52 & 252.51 & 244.40 \\
    Sci-Art & 37.34 & \underline{133.05} & 99.90 & 114.04 & 63.61 \\
    BigCity Aerial & 447.44 & \underline{1137.58} & 918.39 & 1090.98 & 950.52 \\
    \bottomrule
    \end{tabular}
}
\caption{Ablation Study. ``w/o Hier.'' refers to ``Ours'' without hierarchical assignment across machines and then across GPUs within each machine. ``w/o Load Bal.'' disables load balancing in both point assignment and image assignment. ``w/o Patch'' disables splitting images into smaller patches for flexible assignment (§\ref{subsubsec:imageplacement}). Values indicate training throughput (images/s). }
\label{tab:ablation}
\end{table}

Table~\ref{tab:ablation} summarizes the results. We draw two main observations. First, each of the three components contributes meaningfully to performance---removing any of them leads to reduced throughput. In particular, the ``w/o Hier.'' and ``w/o Patch'' variants result in larger degradations than ``w/o Load Bal.''. Specifically, ``w/o Hier.'' is 16\% to 35\% slower than our full system, while ``w/o Patch'' is 10\% to 52\% slower. ``w/o Load Bal.'' shows a smaller impact, with performance reductions ranging from 1\% to 14\%. The relatively modest effect of load balancing stems from the fact that communication time dominates the overall runtime in our setting. The patching technique has an especially high impact for high-resolution scenes such as Sci-Art (6K), where cutting these higher-resolution images into patches gives more potential for flexible assignment. 

Second, despite the removal of any single technique, our system consistently outperforms the baseline much. This is largely due to our core locality-aware distribution framework. It comprises offline METIS bipartite graph partitioning and online linear sum assignment, which together significantly reduce communication overhead compared to random distribution strategies---even without the additional optimizations.

\subsection{System Overheads} 
\label{sec:appendix:system-overheads}

\begin{table}[h]
    \centering
    \resizebox{\columnwidth}{!}{
        \begin{tabular}{c c c c c c c}
            \toprule
            \multirow{2}{*}{\textbf{Scene}} & \multirow{2}{*}{\textbf{Rubble}} & \multirow{2}{*}{\textbf{Ithaca365}} & \multirow{2}{*}{\textbf{Campus}} & \textbf{BigCity} & \multirow{2}{*}{\textbf{Sci-Art}} & \textbf{BigCity} \\
            & & & & \textbf{Street} &  & \textbf{Aerial} \\
            \midrule
            METIS Time (s) & 3.5 & 3.7 & 15.3 & 31.0 & 3.8 & 1.9 \\
            Total Time (s) & 6.1 & 6.9 & 23.9 & 46.9 & 6.0 & 3.4 \\
            \bottomrule
        \end{tabular}
    }
    \caption{Offline point assignment time for each scene in Table~\ref{tab:scenes-main}, including METIS graph partitioning and associated overheads. The resulting partitions are visualized in Figure~\ref{fig:eval_pcd-partition-results}(Appendix \ref{sec:appendix:visualize-partitioning}) and used in the experiments in Figure~\ref{fig:throughput-all-scenes-all-primitives}. } 
    \label{tab:metis-partition-time}
\end{table}

Table~\ref{tab:metis-partition-time} reports offline point assignment times for experiments in Figure~\ref{fig:throughput-all-scenes-all-primitives}. 
This includes METIS graph partitioning and other runtime overheads, with METIS as the primary bottleneck. All steps run on a single process and a single GPU. 
Overheads range from 3.4s to 46.9s---less than 1\% of total training time. This efficiency stems from the graph size after coarsening the point groups \S\ref{subsubsec:pointplacement}, which is reduced to \( \left\lceil \frac{S}{G} \right\rceil + D\) vertices---typically only tens of thousands---making it fast to solve.
We also observe that partitioning time varies by scene. Scenes with more complex topologies and view frustums covering larger areas---i.e. Campus and BigCity Street---take longer to partition. 
Moreover, the end-to-end speedups across scenes (Figure~\ref{fig:throughput-all-scenes-all-primitives}) indicate that online scheduling introduces no significant overhead.

\subsection{Visualization of point cloud partition results} 
\label{sec:appendix:visualize-partitioning}
\begin{figure*}[t]
    \centering

    \begin{subfigure}[b]{0.25\textwidth}
        \includegraphics[width=\linewidth]{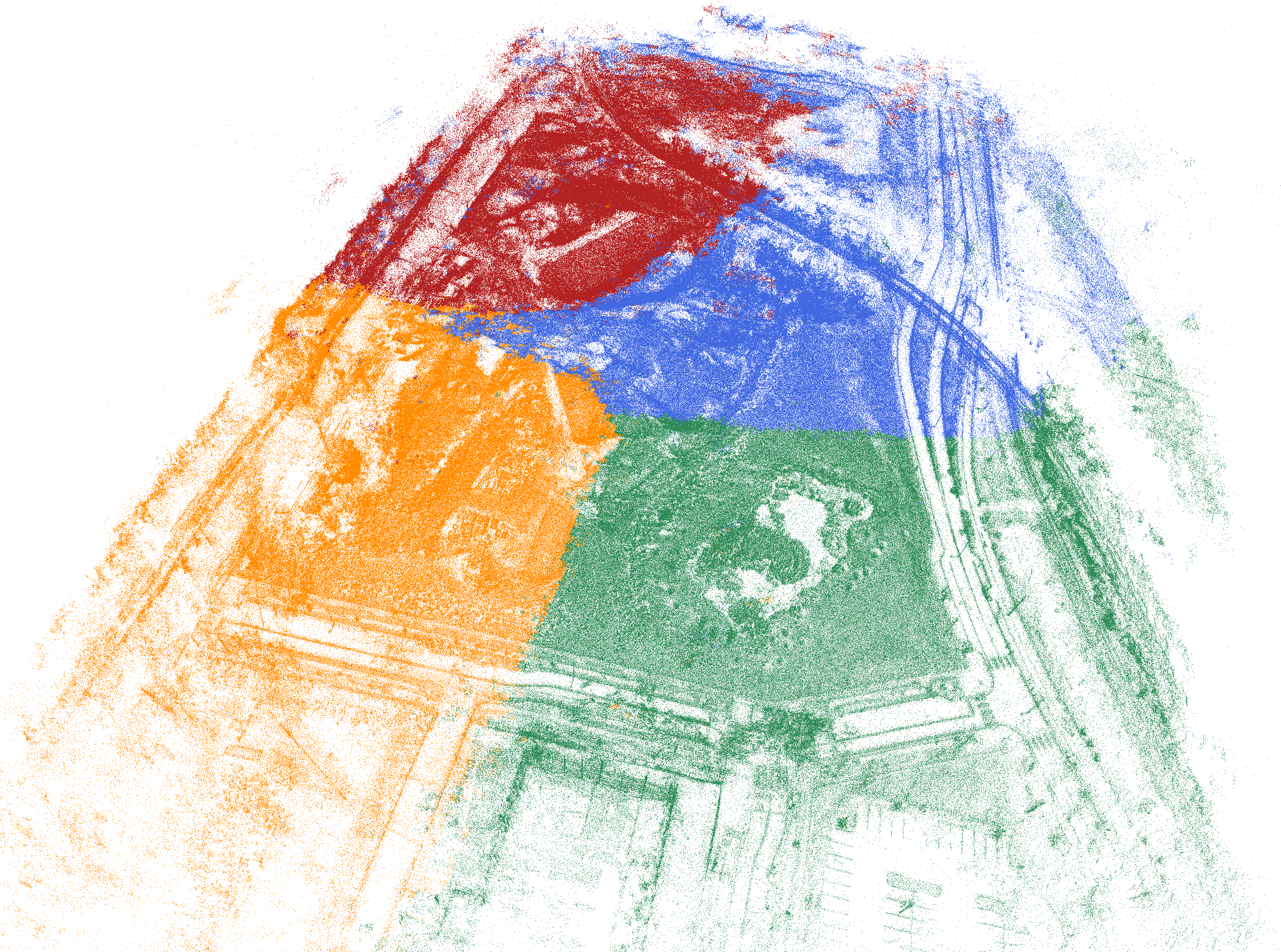}
        \caption{Rubble}
    \end{subfigure}
    \begin{subfigure}[b]{0.32\textwidth}
        \includegraphics[width=\linewidth]{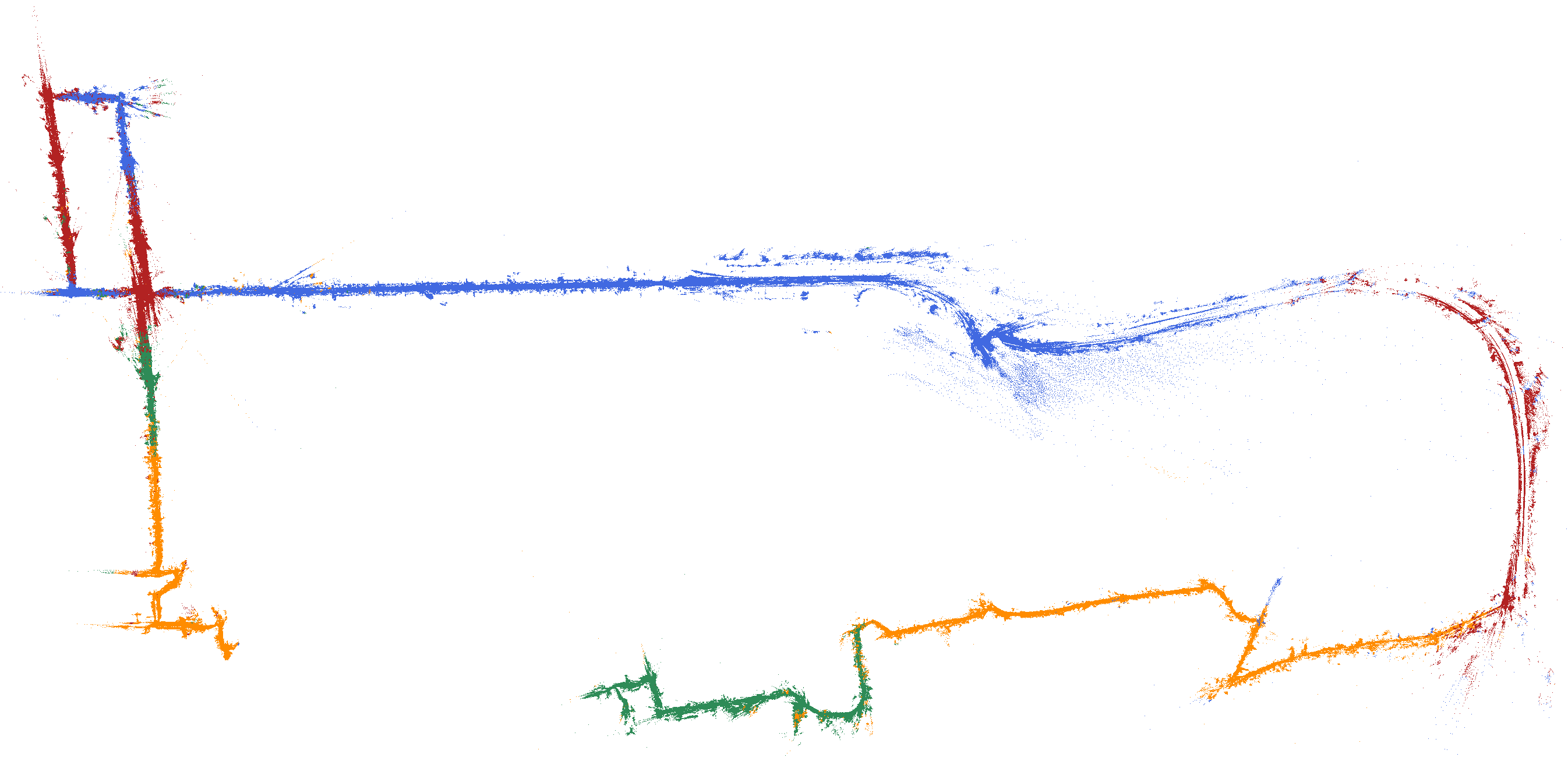}
        \hfill
        \caption{Ithaca365}
    \end{subfigure}
    \begin{subfigure}[b]{0.28\textwidth}
        \includegraphics[width=\linewidth]{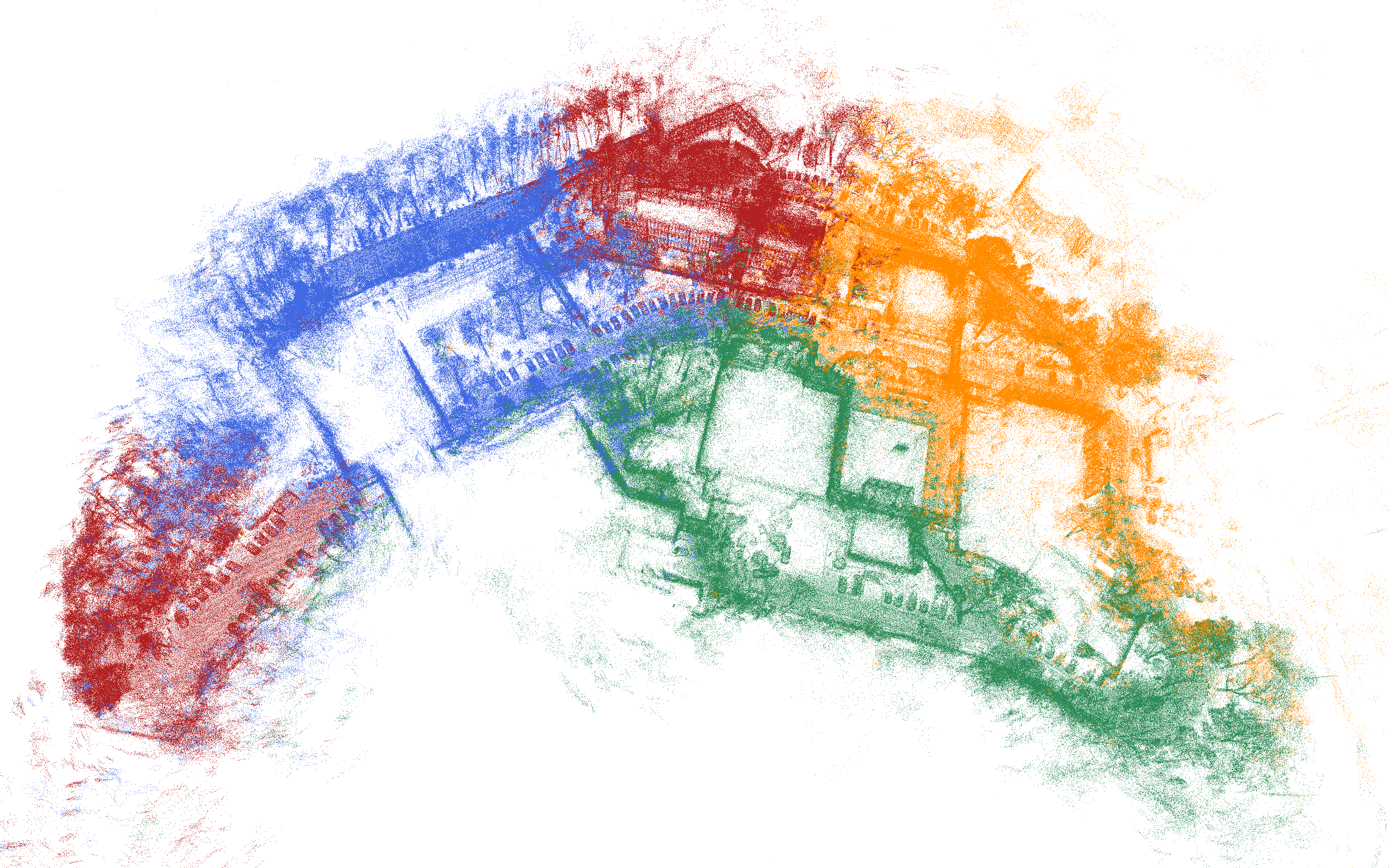}
        \caption{Campus}
    \end{subfigure}

    \vspace{0.5em}  

    \begin{subfigure}[b]{0.3\textwidth}
        \includegraphics[width=\linewidth]{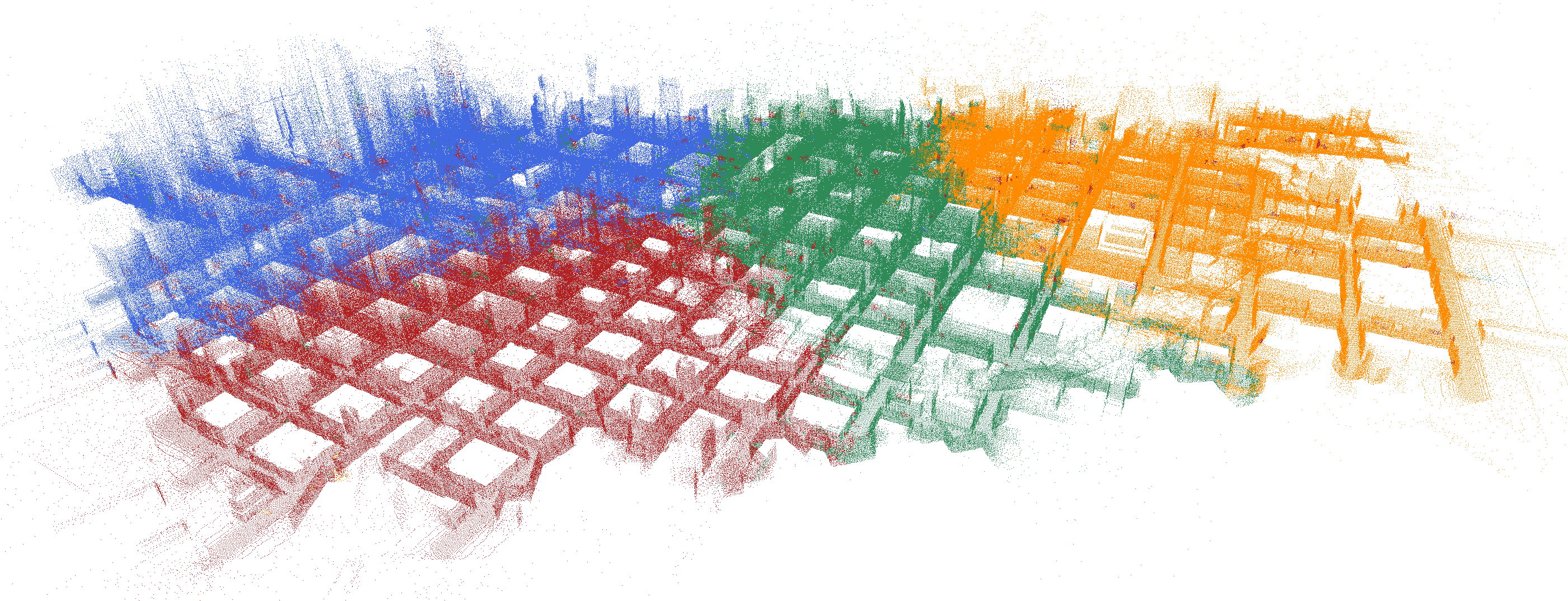}
        \hfill
        \caption{BigCity Street}
    \end{subfigure}
    \begin{subfigure}[b]{0.23\textwidth}
        \includegraphics[width=\linewidth]{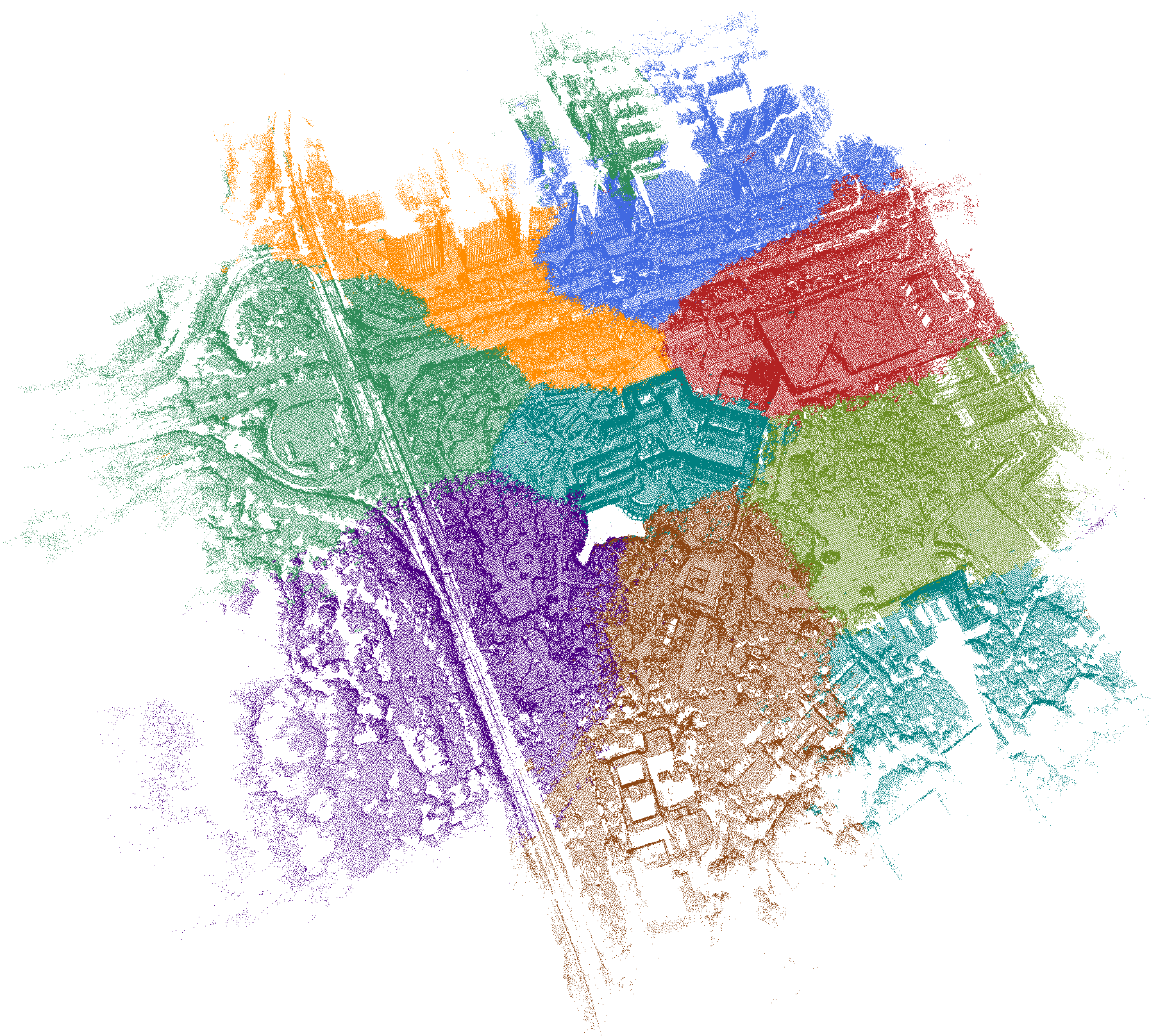}
        \caption{Sci-Art}
    \end{subfigure}
    \begin{subfigure}[b]{0.22\textwidth}
        \includegraphics[width=\linewidth]{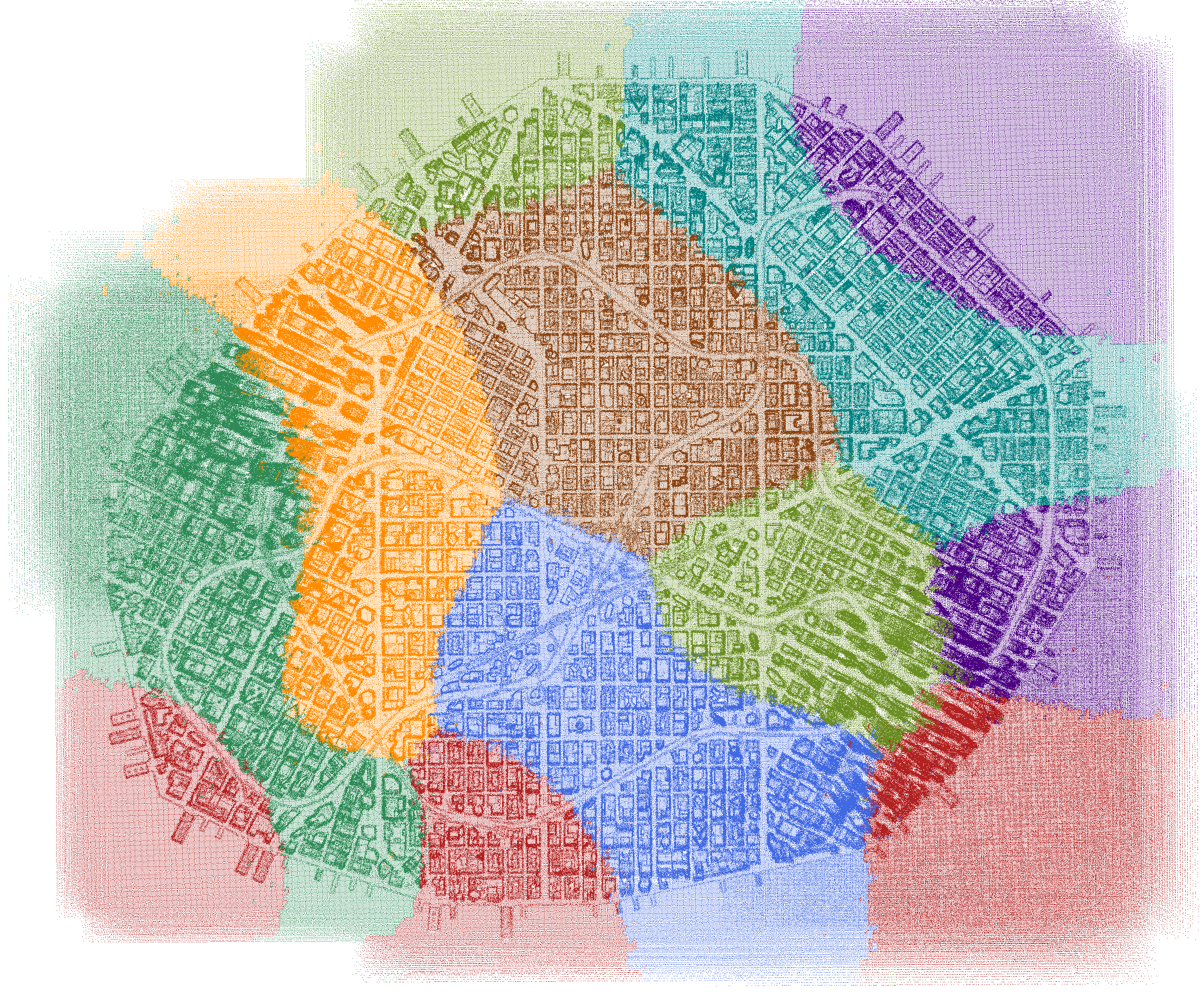}
        \caption{BigCity Aerial}
    \end{subfigure}

    \caption{Point Cloud Partition Results. These are the point clouds for the scenes listed in Table \ref{tab:scenes-main}. Colors indicate the partitioning results, with each color representing the point cloud assigned to a different machine. We can also observe the irregular topology present in the scenes. The point clouds have been downsampled and resized for easier visualization. }
    \label{fig:eval_pcd-partition-results}
\end{figure*}

Figure~\ref{fig:eval_pcd-partition-results} visualizes the point cloud partitions for scenes in Table~\ref{tab:scenes-main} using \name. These visualizations also reveal that the point clouds are irregular and vary significantly across scenes. 

\section{Additional Related Works}
\label{sec:appendix:additional-related-works}

\paragraph{Point Cloud Partitioning Techniques.}
Point cloud partitioning is a key component of our system design. Prior work has explored several strategies: (1) manual heuristics, as used in CityGS~\cite{citygaussian} and VastGS~\cite{vastgaussian}, which divide 3DGS point clouds along coordinate axes; (2) graphics data structures, such as RetinaGS~\cite{retinags}, which employs KD-Tree partitioning; and (3) clustering techniques like k-means, commonly used for general point cloud segmentation~\cite{kmean4pointcloud}. 
However, manual heuristics are often ineffective for irregular point cloud structures. 
Although more general, graphics data structures and clustering methods overlook the role of camera views in partitioning and miss optimization opportunities.

\end{document}